\documentstyle[12pt,dina4,epsfig]{article}
\begin{document}
\baselineskip=20pt
\newcommand{\tri}{\triangle}
\newcommand{\be}{\begin{equation}}
\newcommand{\bel}[1]{\be\label{#1}}
\newcommand{\ee}{\end{equation}}
\newcommand{\bea}{\begin{eqnarray}}
\newcommand{\eea}{\end{eqnarray}}
\newcommand{\beas}{\begin{eqnarray*}}
\newcommand{\eeas}{\end{eqnarray*}}
\newcommand{\ds}{\displaystyle}
\newcommand{\pari}{\scriptscriptstyle\parallel}
%this command corresponds to 'greater than or of the order of':
\newcommand{\goo}{\,\raisebox{-.5ex}{$\stackrel{>}{\scriptstyle\sim}$}\,}
\newcommand{\loo}{\,\raisebox{-.5ex}{$\stackrel{<}{\scriptstyle\sim}$}\,}
\renewcommand\topfraction{0.8}
%=======================================================================
\pagestyle{empty}
\begin{center}

{\Large\bf
Universality of Spectator Fragmentation 
\vspace{0.2cm}

at Relativistic Bombarding Energies}

\vspace*{0.5cm}

\noindent
% enter author
%\author
{
A.~Sch\"uttauf,$^{(1)}$
W.D.~Kunze,$^{(2)}$
A.~W\"orner,$^{(2)}$
M.~Begemann-Blaich,$^{(2)}$ 
Th.~Blaich,$^{(3)}$
D.R.~Bowman,$^{(4)}$\cite{AAA}
R.J.~Charity,$^{(5)}$ 
A.~Cosmo,$^{(6)}$
A.~Ferrero,$^{(7)}$\cite{BBB}
C.K.~Gelbke,$^{(4)}$
C.~Gro\ss,$^{(2)}$
W.C.~Hsi,$^{(4)}$\cite{CCC}
J.~Hubele,$^{(2)}$ 
G.~Imm\'{e},$^{(8)}$ 
I.~Iori,$^{(7)}$ 
J.~Kempter,$^{(1)}$
P.~Kreutz,$^{(1)}$
G.J.~Kunde,$^{(2)}$\cite{DDD} 
V.~Lindenstruth,$^{(2)}$\cite{JJJ}
M.A.~Lisa,$^{(4)}$\cite{JJJ}
W.G.~Lynch,$^{(4)}$
U.~Lynen,$^{(2)}$
M.~Mang,$^{(1)}$
T.~M\"ohlenkamp,$^{(9)}$
A.~Moroni,$^{(7)}$
W.F.J.~M\"uller,$^{(2)}$ 
M.~Neumann,$^{(1)}$
B.~Ocker,$^{(1)}$
C.A.~Ogilvie,$^{(2)}$\cite{FFF}
G.F.~Peaslee,$^{(4)}$\cite{GGG}
J.~Pochodzalla,$^{(2)}$\cite{III}
G.~Raciti,$^{(8)}$
F.~Rosenberger,$^{(2)}$
Th.~Rubehn,$^{(2)}$\cite{JJJ}
H.~Sann,$^{(2)}$
C.~Schwarz,$^{(2)}$
W.~Seidel,$^{(9)}$
V.~Serfling,$^{(1)}$
L.G.~Sobotka,$^{(5)}$
J.~Stroth,$^{(2)}$
L.~Stuttge,$^{(6)}$ 
S.~Tomasevic,$^{(6)}$
W.~Trautmann,$^{(2)}$ 
A.~Trzcinski,$^{(10)}$
M.B.~Tsang,$^{(4)}$ 
A.~Tucholski,$^{(10)}$
G.~Verde,$^{(8)}$
C.W.~Williams,$^{(4)}$
E.~Zude,$^{(2)}$
and B.~Zwieglinski$^{(10)}$
}

\vspace*{0.3cm}

\noindent
% enter address
%\address
{\it
$^{(1)}$Institut f\"ur Kernphysik,
Universit\"at Frankfurt, D-60486 Frankfurt, Germany\\
$^{(2)}$Gesellschaft  f\"ur  Schwerionenforschung, D-64291 Darmstadt, Germany\\
$^{(3)}$Institut f\"ur Kernchemie, Universit\"at Mainz, 
D-55099 Mainz, Germany\\
$^{(4)}$Department of Physics and 
Astronomy and National Superconducting Cyclotron Laboratory, 
Michigan State University, East Lansing, MI 48824, USA\\
$^{(5)}$Department of Chemistry, 
Washington University, St. Louis, MO 63130, USA\\
$^{(6)}$Centre de Recherches Nucl\'{e}aires, F-67037 Strasbourg, France\\
$^{(7)}$ Istituto di Scienze Fisiche, Universit\`{a} degli Studi 
di Milano and I.N.F.N., I-20133 Milano, Italy\\
$^{(8)}$Dipartimento di Fisica dell' Universit\`{a}
and I.N.F.N.,
I-95129 Catania, Italy\\
$^{(9)}$Forschungszentrum Rossendorf, D-01314 Dresden, Germany\\
$^{(10)}$Soltan Institute for Nuclear Studies,
00-681 Warsaw, Hoza 69, Poland
}

\vspace{0.3cm}

%(draft \today)
\end{center}

\begin{tabbing}
Corresponding author: \=                  \kill
Corresponding author: \> W. Trautmann \\
\> GSI, D-64291 Darmstadt, Germany \\
\> e-mail: w.trautmann@gsi.de         \\
\> Tel: x49-6159-71-2774, Fax: x49-6159-71-2989      \\
\end{tabbing}
%omit \newpage when activating this block

%\newpage

{\bf
ABSTRACT}
\vspace{0.3cm}

Multi-fragment decays of $^{129}$Xe, $^{197}$Au, and $^{238}$U projectiles 
in collisions with Be, C, Al, Cu, In, Au, and U targets 
at energies between $E/A$ = 400 MeV and 1000 MeV 
have been studied with the ALADIN forward-spectrometer at SIS.
By adding an array of
84 Si-CsI(Tl) telescopes the solid-angle coverage of the setup was
extended to $\theta_{lab}$ = 16$^{\circ}$. This permitted the complete
detection of fragments from the projectile-spectator source.

The dominant feature of the systematic set of data is the 
$Z_{bound}$ universality that is obeyed by the fragment multiplicities
and correlations. These observables are invariant with
respect to the entrance channel if plotted as a function of
$Z_{bound}$, where $Z_{bound}$ is the sum of the atomic numbers 
$Z_i$ of all projectile fragments with $Z_i \geq$ 2. 
%Except for the lightest targets at the lowest energies,
No significant dependence on the bombarding energy nor on the target
mass is observed. The dependence of the fragment multiplicity on the 
projectile mass follows a linear scaling law. 

The reasons for and the limits of the
observed universality of spectator fragmentation are explored within the
realm of the available data and with model studies. It is found that the
universal properties should persist up to much higher bombarding
energies than explored in this work and that they are
consistent with universal features exhibited by the intranuclear
cascade and statistical multifragmentation models.

\vspace{2cm}


\vspace{0.3cm}


\newpage
\pagestyle{plain}
\setcounter{page}{1}
\pagenumbering{arabic}

\section{Introduction}
\label{Sec_1}

The apparent absence of dynamical dependences is the perhaps most prominent
feature of the multi-fragment decay of excited spectator nuclei.
In the first experiments with the ALADIN spectrometer, performed with
$^{197}$Au beams of $E/A$ = 600 MeV, it has manifested itself as an
invariance of the observed patterns of projectile fragmentation 
with respect to the chosen
target [1-3].
%\cite {ogilvie,hubele1,kreutz}. 
The mean number of projectile fragments produced as well as other
observables characterizing the populated partition space were found to be
the same for all targets, ranging from carbon to lead, if they were 
plotted as a function
of $Z_{bound}$. The quantity $Z_{bound}$ is defined
as the sum of the atomic numbers $Z_i$ of all projectile fragments
with $Z_i \geq$ 2. It represents the charge of the spectator system
reduced by the number of hydrogen isotopes emitted during its decay.

It is easy to argue that $Z_{bound}$ may, rather
directly, reflect the energy transfer to the excited spectator system.
The removal of nucleons in the initial excitation stage of the reaction
and the release of hydrogen isotopes during the subsequent 
deexcitation are both correlated with the energy transfer.
Larger energy transfers correspond to smaller 
values of $Z_{bound}$. The observed target invariance 
hence suggests that the energy transfer to the projectile spectator 
is the primary quantity governing its decay.

These characteristics represent, at least, 
an indication that statistical equilibrium is attained prior to 
the fragmentation stages of the reaction. 
In fact, statistical models were found to be quite 
successful in describing the measured fragment yields and
correlations, provided that emission from an expanded breakup
state was assumed [4-9].
%\cite{hubele2,botv1,barz1,baoanli,konopka,zheng}.
Very recently, a near perfect description of the experimental 
charge correlations measured for the reaction 
$^{197}$Au on Cu at $E/A$ = 600 MeV, including
their dispersions around the mean behavior, was achieved with the
statistical multifragmentation model \cite{botv2}. This comparison was
made on an absolute scale, apart from an overall normalization
constant which relates the number of model events to the measured cross
section. 
%A continuous distribution of excited and 
%equilibrated residual nuclei, used as the input for the calculations, was
%derived by searching for the best fit to the data. The resulting distribution 
%is characterized by a correlation between a decreasing mass number $A$ of the
%residue and an increasing excitation energy $E_x/A$.
 
Intriguingly, the fragment-charge correlations 
from these first experiments were also reproduced to high 
accuracy with a variety of other models, such as 
site-bond-percolation \cite{kreutz}, classical-cluster formation 
\cite{garcia}, fragmentation-inactivation binary \cite{botet},
and restructured-aggregation models \cite{leray}.
Some of these models are of a predominantly 
mathematical nature, have very few
parameters, and do not contain any of the nuclear
physics input on which the statistical models are based. Some exhibit
critical behavior in the limit of an infinite number of constituents.
It will be an interesting task to identify
the underlying symmetries, apparently common to all of these
models, and to study their relation to the universal properties
of the fragment decay of excited
nuclei \cite{richert}. 

The question of equilibration in the multi-fragment decay 
of excited spectator systems is of highest interest.
Multifragmentation has been considered a manifestation of the 
liquid-gas phase transition
in finite systems. The pursuit of this challenging problem
(see, e.g. [15-17]
%\cite{moretto,gilkes,pocho1} 
and references given therein)
will profit from the study of equilibrated systems 
that may be characterized by a few global
variables such as mass, excitation energy, and temperature. 
The correlation of energy transfer and $Z_{bound}$ opens the
possibility to experimentally control the former quantity by selecting
on the latter \cite{pocho1}.

The present work was performed in order to further
establish the validity of the observed universality 
and to search for its possible limits. 
The multi-fragment decay of heavy projectiles 
was explored over a
wider range of projectiles and targets and within the regime of 
relativistic bombarding energies between 400 and 1000 MeV per nucleon.
For these new experiments, the ALADIN spectrometer \cite{hubele1}
has been upgraded by installing a new tracking detector TP-MUSIC III, 
by enlarging the time-of-flight wall behind it, and by mounting 
a new Si-CsI(Tl) hodoscope which provided good coverage of the solid-angle 
adjacent to the entrance of the spectrometer. 
In the data analysis, a consistent definition of the 
projectile-spectator source was adopted prior to comparing results
from different reactions. 

It was found that, within the realm of the present investigation,
the universality of spectator fragmentation holds to very high accuracy. 
This includes the fragmentation of different projectiles 
if a linear scaling in proportion to their masses is applied to the data.
In the following sections the experimental method and the obtained 
results will be presented in detail. The subsequent discussion will include
comparisons to other data, some taken at much higher bombarding energies,
and a search for related universal features within the intranuclear cascade 
and statistical multifragmentation models.

\section{Experimental method}
\label{Sec_2}

The experiments were performed at the heavy-ion 
synchrotron SIS of the GSI Darmstadt. 
Beams of $^{129}$Xe at $E/A$ = 600 MeV incident energy, 
of $^{197}$Au at $E/A$ = 400, 600, 800, and 1000 MeV,
and of $^{238}$U at $E/A$ = 600 and 1000 MeV were used.
Their intensities were
of the order of a few thousand
particles per second over spill lengths of several seconds. 
The areal densities
of the targets made from Be, C, Al, Cu, In, Au and U were 190,
200, 390, 420, 800, 480, and 480 mg/cm$^2$, respectively, and
corresponded to
interaction probabilities between $\approx$ 0.7\%
and 3.9\%.
Complete sets of targets were bombarded with the
$^{197}$Au beams (Be through Au) and with the $^{238}$U beams 
(Be through U) at energies $E/A$ = 600 and 1000 MeV. 
Only selected targets were used with 
the $^{129}$Xe beam and with the $^{197}$Au beams of $E/A$ = 400
and 800 MeV.

\begin{figure}[p]
     \centerline{
     \epsfig{file=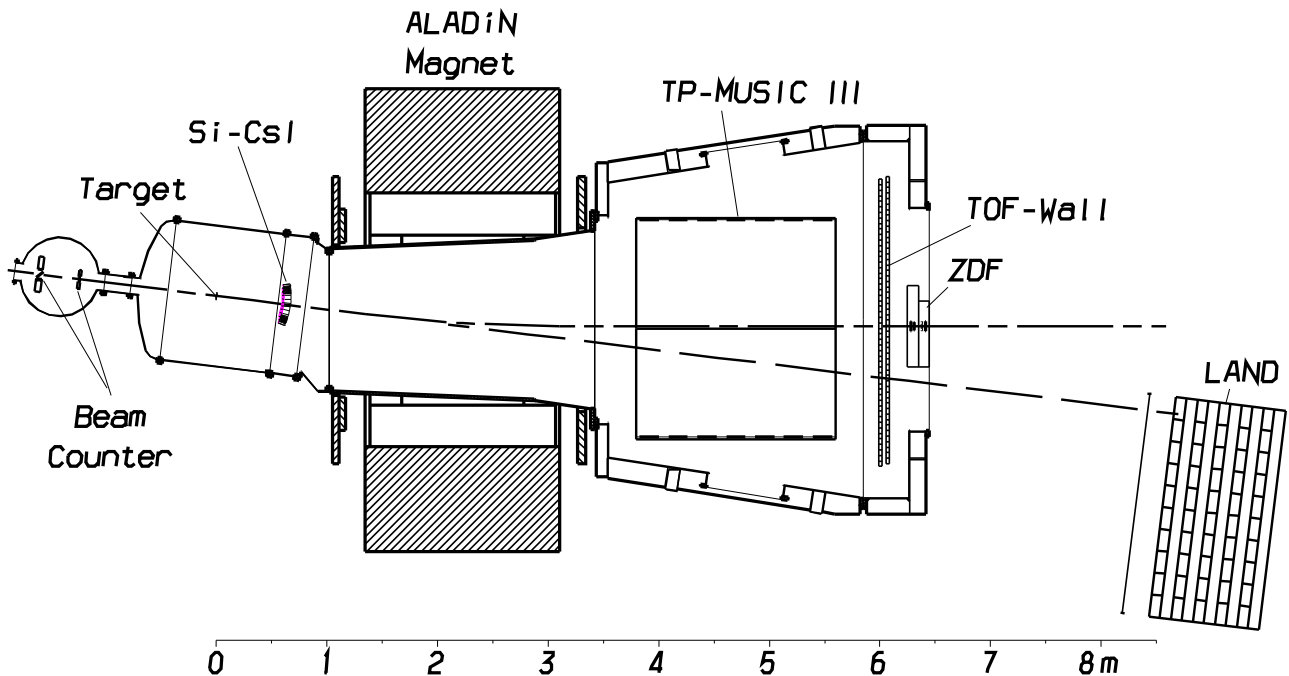,width=17.cm}}
     \begin{center}
     \parbox{15.cm}{\caption[setup s114]{\em
Cross sectional view of the ALADIN facility. The beam enters from the
left and is monitored by two beam detectors before reaching the target.
Projectile fragments entering into the acceptance of the magnet are 
tracked and identified in the TP-MUSIC III detector and in the 
time-of-flight (TOF) wall. The Central Plastic detector covers the hole 
in the TOF wall at the exit for the beam.
Fragments and particles emitted in forward
directions outside the magnet acceptance and up to 
$\theta_{lab} = 16^{\circ}$ are detected in the Si-CsI array.
Neutrons emitted in directions close to 0$^{\circ}$ are detected with the
large-area neutron detector (LAND).
The dashed line indicates the direction
of the incident beam. The dash-dotted line represents the trajectory of
beam particles after they were deflected by an angle of 7.3$^{\circ}$.
The Miniball/Miniwall detector system used at $E/A$ = 400 MeV is not shown. 
     \label{setup}
     }}
     \end{center}
\end{figure}

A schematic layout of the experimental setup is shown in fig.~\ref{setup}.
For each beam particle, its arrival time and its position in the
plane perpendicular to the beam direction were measured
upstream of the target with two thin plastic scintillators. Their
effective thicknesses were 110 $\mu$m and 50 $\mu$m. 
The geometric acceptance of the spectrometer of 
$\theta_{lab} \approx \pm 9.2^{\circ}$ horizontally 
and $\pm 4.3^{\circ}$ vertically
was matched by the dimensions of the new multiple sampling ionization
chamber TP-MUSIC III and by the extended time-of-flight (TOF) wall. 
At $E/A$ = 1000 MeV, these detector systems permitted the detection of
close to 100\% of all 
projectile fragments with atomic number $Z \ge$ 2. At the lower 
bombarding energies, the angular distributions of some 
lighter fragments extended beyond the acceptance of the spectrometer
but stayed within the acceptance of the Si-CsI(Tl) hodoscope array
that surrounded the entrance to the field gap of the magnet.
For very light fragments, up to $Z \approx$ 4,
the emissions from the projectile and
mid-rapidity sources are not clearly separated in momentum space.
Therefore, a consistent definition of the projectile spectator source 
had to be adopted for the data analysis. This 
will be described in section 3.1.

The atomic numbers $Z$ and the velocities 
of nuclear fragments were determined with the TOF wall, located at the end
of the ALADIN spectrometer and extending over 2.4~m in horizontal and
1.0~m in vertical directions. It consisted of two layers of vertically
mounted scintillator strips of 2.5 cm width and 1.0 cm thickness,
viewed by photomultiplier tubes at both ends \cite{hubele1}. 
The two layers were offset by half a width with respect to 
each other. The primary beam was directed through a hole of
4.8 x 6.0 cm$^2$ cut into the middle sections of the central slats.
The upper and lower halves of these central slats were optically connected
by a hollow light guide built from aluminized mylar foil.
The discriminator threshold was set to
be below $Z$ = 2 particles. 
For the time-of-flight calibration, primary beams of reduced intensity 
were swept across the wall with the Aladin magnet. By inducing 
fragmentation reactions in a thick aluminum target positioned 
immediately in front of the wall the response
to fragments of different $Z$ was determined.
A resolution of about 400 ps and 180 ps (FWHM) for fragments of $Z$ = 2
and $Z \approx Z_P$, respectively, was achieved ($Z_P$ denotes the atomic 
number of the projectile).
These values include the systematic uncertainty of the calibration.
The intrinsic time-of-flight resolution measured for beam
particles was 120 ps.

\begin{figure}[p]
     \centerline{
     \epsfig{file=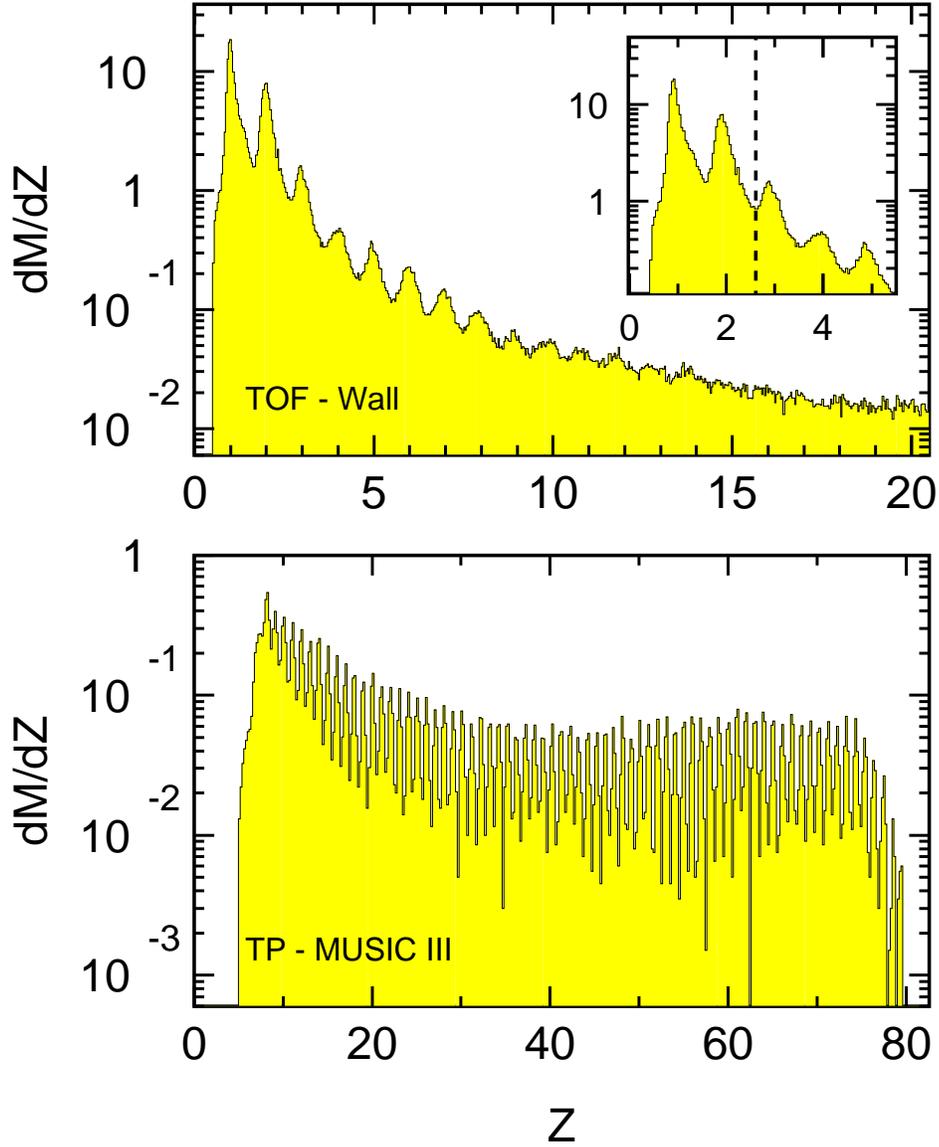,height=16.cm}}
     \begin{center}
     \parbox{15.cm}{\caption[TOF MUSIC Zspek]{\em
$Z$-identification spectra measured with the TOF wall
(top) and the TP-MUSIC (bottom) for the reaction
$^{197}$Au on $^{197}$Au at $E/A$ = 600 MeV.
The $Z$ information from the TP-MUSIC
was used to calibrate the response of the TOF wall in the
region $Z >$ 15.
Note that
the element yield at $Z >$ 65 is affected by the experimental trigger.\\
Insert: Low-$Z$ part of the TOF-wall spectrum.
The dashed line indicates the equivalent sharp cut which 
was used for selecting fragments with $Z \ge$ 3.
     \label{zspectra}
     }}
     \end{center}
\end{figure}

The algorithm used for fragment identification in the TOF-wall analysis
took into account that fragments may pass into the narrow gap between
adjacent slats in one of the two layers of the wall and that heavy
fragments are accompanied by a considerable number of $\delta$-rays.
The identification spectrum for the TOF-wall detectors
was obtained by projecting along the ridges
of constant $Z$ in two-dimensional maps of the measured pulse height
versus time of flight. 
Elements with $Z \leq$ 15 were resolved individually.
For heavier fragments the resolution
assumed values of up to $\Delta Z \approx$ 1.5 (FWHM). 
For this analysis, the TP-MUSIC III
detector, capable of identifying the individual elements for
$Z \geq$ 8 with a resolution between $\Delta Z \approx$ 0.8 
(FWHM, $Z \approx$ 20) and $\Delta Z \approx$ 0.4 ($Z \approx$ 60), 
served to calibrate the charge response of the TOF wall.
The main purpose of the TP-MUSIC, in these experiments, was to provide the 
tracking information for other analyses that involved the fragment momenta. 
The dynamic range for tracking had been extended down to $Z$ = 2 by using
gas amplification over part of the length of the detector which
permitted the measurement of isotopic yield ratios of light fragments
\cite{pocho1}.
The high charge resolution of the TP-MUSIC was essential for the 
analysis of fission decays in experiments with the uranium 
beams [18-20].
%\cite{rubehn1,rubehn2,rubehn3}.

The $Z$ resolution achieved with the two detector systems after
off-line calibration is demonstrated in fig.~\ref{zspectra}. 
A good separation of $Z$ = 2 and 3 fragments was
crucial for the reliable determination of the multiplicity 
of intermediate-mass
fragments with $Z~\ge$~3. Therefore, the $Z$~=~2 and 3
distributions in the $Z$-identification spectrum were fitted individually
and, after evaluation of their overlapping tails, the position of 
an equivalent sharp $Z$-cut was determined (fig. 2, insert).

\begin{figure}[htb]
     \centerline{
     \epsfig{file=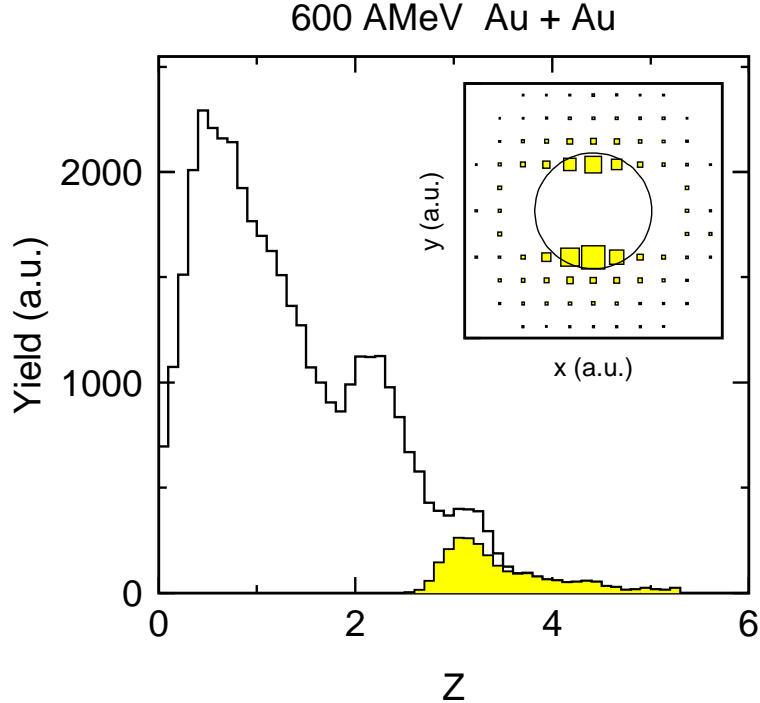,height=10.cm}}
     \begin{center}
     \parbox{16.cm}{\caption[Hodo id]{\em
$Z$-identification spectrum, obtained from a weighted sum of the two
energy-loss measurements with the Si-CsI(Tl) telescopes of the 
hodoscope for the reaction
$^{197}$Au on $^{197}$Au at $E/A$ = 600 MeV.
Intermediate-mass fragments 
of $Z \ge$ 3, as identified in the two-dimensional energy-loss 
spectrum, are represented by the dark-shaded distribution.\\
Insert:
Distribution of intermediate-mass fragments across the solid angle
subtended by the hodoscope detectors. Each square represents a
telescope module with its area being proportional to the detected
fragment yield.
The circle is meant to illustrate that six hodoscope detectors are within
the angular limit of the spectator source at this incident energy.
     \label{hodoid}
     }}
     \end{center}
\end{figure}

The Si-CsI(Tl) hodoscope 
was positioned 60 cm downstream from the target and
covered the solid angle surrounding the spectrometer acceptance up to
angles $\theta_{lab} \approx 16^{\circ}$.
Its 84 telescope modules were mounted in close
geometry. 
Each detector consisted of a 300-$\mu$m Si detector followed
by a 6-cm long CsI(Tl) detector which was viewed by a 1-cm$^2$ 
photodiode from the end face. The active area of each detector was
30 x 30 mm$^2$; the solid-angle coverage of the hodoscope
with respect to the subtended solid angle, i.e. not counting the central 
rectangular opening,
amounted to 85\%. Light fragments with approximately beam velocity
were not stopped by the telescopes. For these particles, the 
identification had to be based on the two-fold measurement of
their energy loss. The peaks caused by fast hydrogen, helium, and
lithium ions are clearly visible in the 
identification spectrum obtained from a weighted 
sum of the two $\Delta E$ measurements with the hodoscope
detectors (fig.~\ref{hodoid}).

For the actual analysis, gates were set in
the two-dimensional
$\Delta E_1$ versus $\Delta E_2$ spectra which permitted the identification
of stopped fragments and a more efficient
suppression of background. The yield 
of fragments with $Z \geq$ 3, selected in this way,
is given by the dark-shaded distribution shown in fig.~\ref{hodoid}. 
The insert shows the
distribution of these fragments, mostly lithium, across the solid
angle subtended by the hodoscope for the reaction 
$^{197}$Au on $^{197}$Au at $E/A$ = 600 MeV.
Each square represents a detector
element, with its area being proportional to the number of detected fragments.
The circle indicates the angular limit up to which fragments were
considered as belonging to the spectator source at 
this incident energy (see section 3.1).

For the runs at 600, 800, and 1000 MeV per nucleon, a 5-mm thick central
plastic detector with light-fiber readout was installed
behind the central hole of the TOF wall. It served for the detection
of the primary beam particles
and of projectile fragments emitted along directions 
close to that of the beam, and for their identification according to the 
measured energy-loss signal.
The large-area neutron detector LAND
was positioned close to zero degrees with respect to the incident beam
direction and operated in a calorimetric mode. First results obtained from
the measurement of coincident neutrons with
LAND have been reported elsewhere \cite{pocho1,bormio_95}.
The experiment at 400 MeV per nucleon was performed with the
Miniball/Miniwall \cite{desouza} installed around the target. 
This detector system covered the angular range 
from $\theta_{lab}$ = 14.5$^{\circ}$
to $\theta_{lab}$ = 160$^{\circ}$ and, together with the 
Si-CsI(Tl) hodoscope, provided an efficient 4-$\pi$ multi-particle
detection capability.
Results from the coincident detection of fragments from the target 
and projectile spectators and from the mid-rapidity source are 
given in ref. \cite{tsang}. For the present study of the projectile
spectator, the increased coverage of the mid-rapidity emission
by the Miniball/Miniwall detectors permitted a good event selection 
according to multiplicity. This was particularly useful as the central 
plastic detector was not in operation at this energy. Peripheral 
events with a small apparent $Z_{bound}$, due to the escape of
a heavy projectile residue through the central hole of the TOF wall, were 
identified by their small associated multiplicity.

The on-line trigger condition consisted of the logical product of
the requirements of a beam particle in the start
detector, no fragment with $Z$ close to that of the beam
in the central plastic detector or in the 
central part of the TOF wall,
and the detection of at least one particle with the Si-CsI(Tl) 
hodoscope. 
It had the effect of suppressing the most peripheral
interactions except those leading to binary fission.
For the present study of multi-fragment production, the peripheral
fission events which may have large cross sections, in
particular for the case of $^{238}$U projectiles \cite{rubehn1},
were suppressed off-line.
In the experiment with the $^{197}$Au beam of 400 MeV per nucleon, 
a beam particle in the start
detector and a minimum of one particle detected with the
Miniball/Miniwall or the Si-CsI(Tl) hodoscope were required.
Scaled down events with less restrictive trigger
conditions, including beam events triggered only by the beam detectors,
were recorded for normalization.

Absolute cross sections were determined by normalizing the measured
event rate with respect to the thickness of the target and the 
rate of incoming beam particles. The error of the normalization,
dominated by the uncertainty of the areal density of the targets, is between
1\% and 5\%. The uncertainties of the
measured fragment multiplicities and correlations 
will be discussed in section 3.3.

\section{Experimental results}
\label{Sec_3}

{\bf 3.1 Spectator source}
\vspace{0.2cm}

The ALADIN spectrometer has been designed for optimum acceptance
of projectile fragments which are emitted in forward directions.
However, the effect of forward focussing by the reaction kinematics
is a function of the projectile
velocity and changes with changing bombarding energy.
We have therefore studied the extension of the projectile-spectator
source in rapidity $y$ and in transverse momentum $p_t$, both experimentally
and with model calculations.

\begin{figure}[p]
     \centerline{
     \epsfig{file=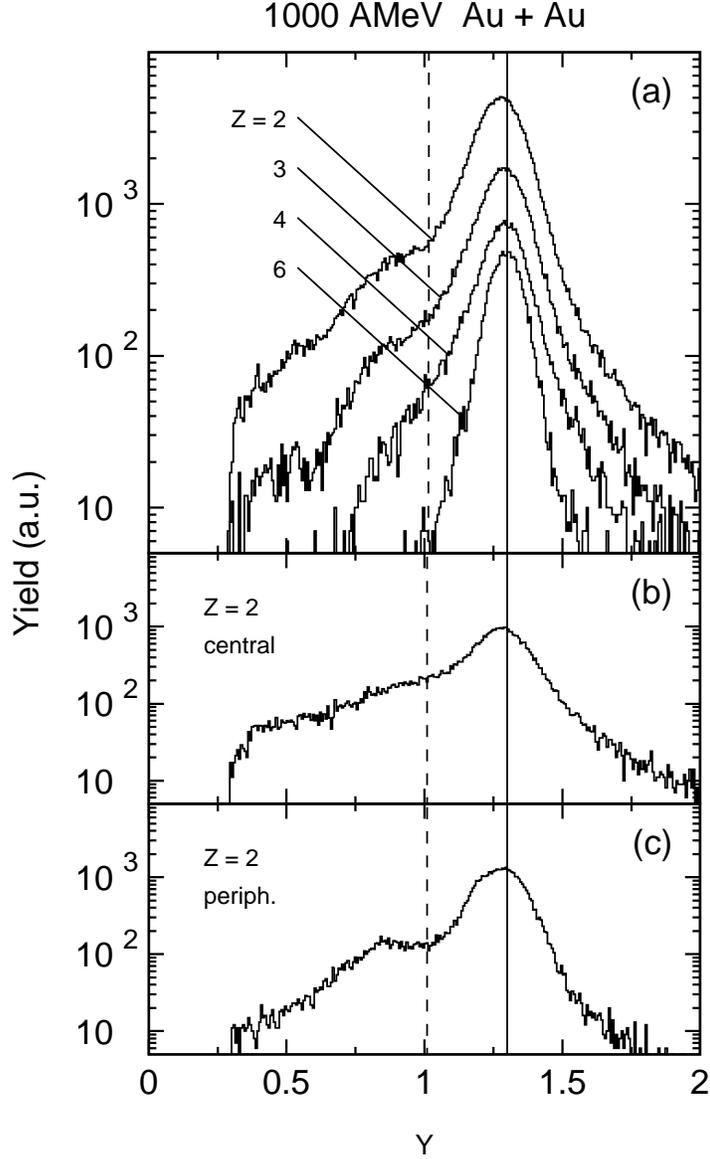,height=16.cm}}
     \begin{center}
     \parbox{15.cm}{\caption[Rapid 1000]{\em
(a): Rapidity spectra measured in the reaction $^{197}$Au on $^{197}$Au at
$E/A$ = 1000 MeV for fragments with $Z$ = 2, 3, 4, and 6.
The solid and dashed lines indicate the measured most probable 
rapidity $y =$ 1.32
of the light fragments and the condition $y \geq 0.75\cdot y_P$ 
adopted for fragments from the projectile spectator, respectively.\\
(b): Rapidity spectra of helium fragments, measured in central 
collisions ($Z_{bound} \leq$ 30)
for the same reaction.\\
(c): Same as (b) for peripheral collisions ($Z_{bound} \geq$ 50).
     \label{rapid}
     }}
     \end{center}
\end{figure}

Rapidity spectra of light fragments detected with the TOF wall
in the reaction $^{197}$Au on $^{197}$Au at
$E/A$ = 1000 MeV are shown in fig.~\ref{rapid}. The distributions are
concentrated around a rapidity value very close to the beam
rapidity and become increasingly narrower with increasing
mass of the fragment. The deviation of the most probable
rapidity $y$ = 1.32 of light fragments (full line in fig.~\ref{rapid})
from the rapidity $y_P$ = 1.35 of the beam, after it has passed 
through half of the Au target,
is within the uncertainty of the absolute time-of-flight calibration.
For the lighter fragments, the distributions are wider 
and extend into the mid-rapidity region. The widths and
shapes of the distributions also depend on the impact parameter, as
demonstrated for helium fragments in the two lower panels of fig.~\ref{rapid}.
Both the dispersion around the projectile rapidity and the 
relative intensity at mid-rapidity increase with increasing centrality.

The bump in the peripheral He spectrum, located at a rapidity $y$ between
0.8 and 0.9, originates from mid-rapidity emission. This was confirmed
by simulating the emission with two Maxwellian sources
centered at projectile rapidity and at mid-rapidity. 
The restriction of the acceptance to forward angles causes the maximum of
the observed mid-rapidity yield to appear at this somewhat larger rapidity.
For this reason, the two sources can only be discerned 
in the peripheral collisions with lower emission temperatures. 
In more central collisions, the mid-rapidity source no longer produces
an identifiable bump in the rapidity spectrum.
Indications of a mid-rapidity source are only observed for light fragments
up to $Z \approx$ 4 (fig.~\ref{rapid}, top), in agreement with the rapid drop
of the element yields in central collisions at relativistic 
bombarding energies \cite{reisdorf}. 

Guided by these observations, 
we chose a lower limit in rapidity of
$y \geq 0.75\cdot y_{P}$ where $y_{P}$ is the
projectile rapidity, so as to define the spectator source.
This condition was applied to fragments
detected with the TOF wall. For those detected with the hodoscope,
for which an equivalent time-of-flight information does not exist,
angular limits were set with the same purpose
of selecting the spectator source. They are motivated and described
below.

\begin{figure}[tb]
     \centerline{
     \epsfig{file=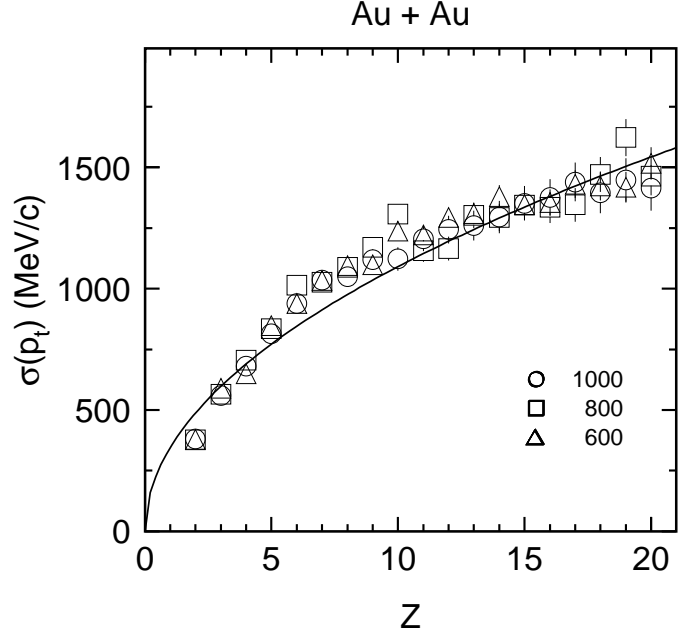,height=9.cm}}
     \begin{center}
     \parbox{15.cm}{\caption[Transversal momenta]{\em
Widths of the transverse-momentum distributions
$\sigma (p_t)$ as a function
of $Z$ for the reactions  $^{197}$Au on $^{197}$Au at
$E/A$ = 600, 800, and 1000 MeV and for 20 $\le Z_{bound} \le$ 60.
The line is proportional to $\sqrt{Z}$.
     \label{transm}
     }}
     \end{center}
\end{figure}

Transverse momentum distributions were constructed from the 
vertical position distributions of fragments measured with the
TOF wall. The vertical coordinate is perpendicular to the bending plane 
of the magnet. In fig.~\ref{transm} we show the widths
\begin{equation}
\sigma (p_t) \approx \sqrt{2} \cdot 
\sigma (r_y/L) \cdot A \cdot p_{P}/A_{P}
\label{EQ1}
\end{equation}
as a function
of $Z$ for the reactions  $^{197}$Au on $^{197}$Au at
$E/A$ = 600, 800, and 1000 MeV.
Here $r_y/L$ is the ratio of the vertical position $r_y$ of a fragment to the
length $L$ of the flight path and $\sigma (r_y/L)$ is the width of the
distribution as obtained by Gaussian fitting. The length $L$ was
approximated by using the
nominal trajectory, starting at the target with 
$\theta_{lab}$ = 0$^{\circ}$,
to the particular slat of the TOF wall that was hit.
$A$ is the mass number assumed for fragments with a given $Z$,
$p_{P}$ and $A_{P}$ are the momentum and the mass number of the 
projectile. We have used 
$A$ = 2$\cdot Z$ for $Z \le$ 10 and
the EPAX parameterization \cite{suemmerer} for heavier fragments.

The widths $\sigma (p_t)$ rise approximately in proportion to 
$\sqrt{Z}$, i.e. the mean transverse energies are independent
of Z which is expected for emission from an equilibrated 
source (see section 3.5).
It is, furthermore, evident that the transverse momentum 
distributions of the emitted projectile
fragments do virtually not change with the bombarding energy.
There are no indications of any dynamical contributions to the fragment 
velocities in the moving frame which could be related to the incident 
energy. Therefore, in order to select the spectator source,
upper limits of the laboratory angle were imposed on the 
hodoscope data that were inversely
proportional to the projectile momentum per nucleon. 
The same limits were used for all three projectiles because
the dependence of the $p_t$ distributions on the projectile mass
was found to be insignificant for the light fragments.

At 1000 MeV per nucleon, the adopted angular limit
coincides with the vertical acceptance of the ALADIN magnet 
of $\theta_{lab} = \pm 4.3^{\circ}$. 
For the helium isotopes,
this corresponds to the condition $|p_y| \le 2\sigma (p_y)$ where
$\sigma (p_y)$ is the width of the transverse momentum distribution
in vertical direction.
At the lower incident energies 800, 600, and 400 MeV per nucleon,
the resulting angular limits are
$\theta_{lab} = 5.0^{\circ}$, $6.0^{\circ}$, and $7.7^{\circ}$, respectively.
Consequently, the fragment yields detected with the corresponding 
central detectors of the Si-CsI(Tl) telescope array were counted as 
belonging to the projectile spectator. At 600 MeV per nucleon, this
included six detectors, as illustrated in fig.~\ref{hodoid} (insert).
They contributed about
10\% of the fragment yield at this energy, 
mostly very light fragments with $Z$ = 3 and above.\\

{\bf 3.2 Charge correlations for $^{197}$Au + $^{197}$Au}
\vspace{0.2cm}

\begin{figure}[tb]
     \centerline{
     \epsfig{file=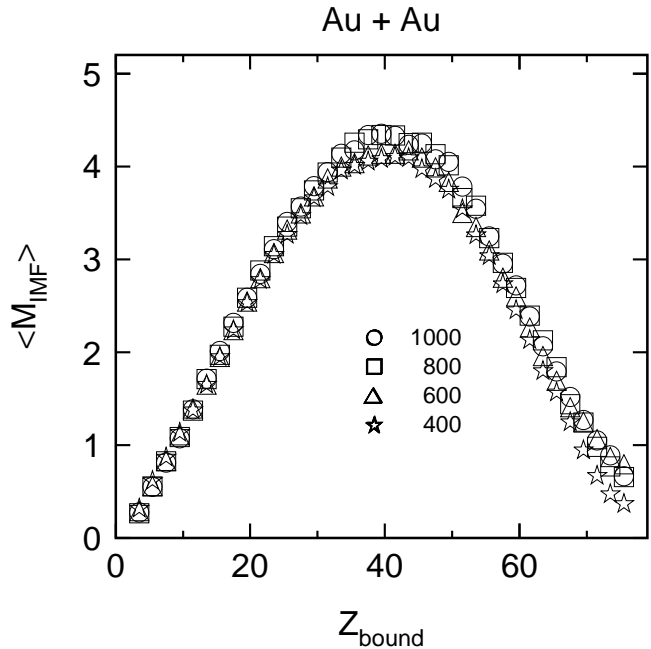,height=9.cm}}
     \begin{center}
     \parbox{15.cm}{\caption[Universalaty Eeer]{\em
Mean multiplicity of intermediate-mass
fragments $\langle M_{IMF} \rangle$ 
as a function of $Z_{bound}$ for the reaction
$^{197}$Au on $^{197}$Au at $E/A$ = 400, 600, 800, and 1000 MeV.
     \label{uniener}
     }}
     \end{center}
\end{figure}

With the definitions of the spectator source in rapidity and
angle, as given in the previous section, the fragmentation 
patterns in $^{197}$Au on $^{197}$Au collisions were studied.
Figure~\ref{uniener} shows the correlation between the mean multiplicity 
of intermediate-mass fragments $\langle M_{IMF} \rangle$ and 
$Z_{bound}$ for the four bombarding energies. 
Here intermediate-mass fragments (IMF's) were selected according to 
the definition 3 $\le Z \le$ 30. The familiar rise and
fall of the fragment production is seen to be independent of 
the projectile energy within the experimental accuracy. 
The invariance with respect to the target, as observed earlier [1-3]
%\cite{ogilvie,hubele1,kreutz} 
and discussed further in section 3.4, 
is thus complemented by an invariance over 
the range of bombarding energies studied in the present experiment. 
At $Z_{bound} \approx$ 40 the mean multiplicity 
$\langle M_{IMF} \rangle$ reaches its maximum of 4 to 4.5.
Due to the enlargement of the spectrometer 
acceptance and the chosen source definition, this value
exceeds the one
reported earlier [1-4]
%\cite{ogilvie,hubele1,kreutz,hubele2} 
by one half to one unit.
For the same reason, the position of the maximum is shifted to a 
slightly larger value of $Z_{bound}$.

\begin{figure}[tb]
     \centerline{
     \epsfig{file=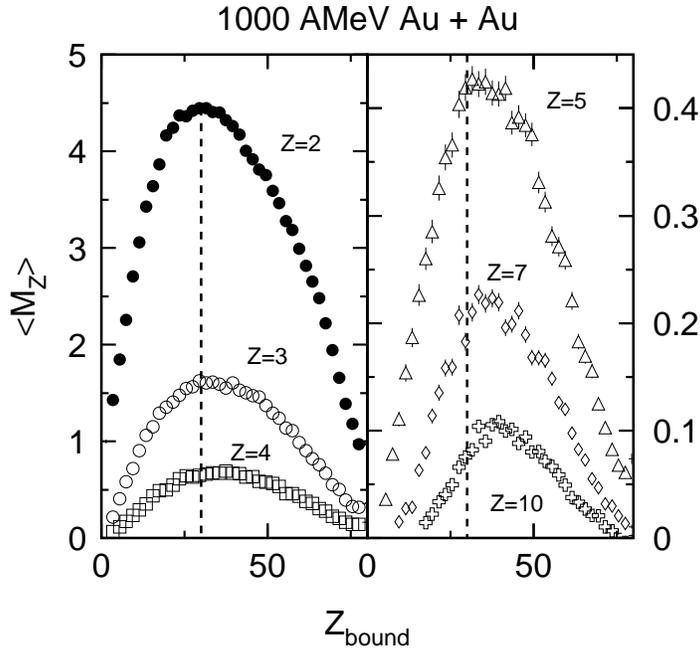,height=9.cm}}
     \begin{center}
     \parbox{15.cm}{\caption[Z dependence]{\em
Mean values of the multiplicities $\langle M_Z \rangle$ for selected
elements as a function of 
$Z_{bound}$ for the reaction
$^{197}$Au on $^{197}$Au at $E/A$ = 1000 MeV.
The dashed lines mark the value $Z_{bound}$ = 30 where 
$\langle M_2 \rangle$ assumes its maximum.
     \label{zdepen}
     }}
     \end{center}
\end{figure}

The definition of the spectator source is most crucial for the 
multiplicities of the lightest
projectile fragments such as helium, lithium, and beryllium nuclei.
Heavier fragments are more strongly localized in velocity space 
(cf. figs.~\ref{rapid} and~\ref{transm}). 
Figure~\ref{zdepen} shows the mean multiplicities 
$\langle M_Z \rangle$ for several light
elements as a function of $Z_{bound}$ for the incident energy
1000 MeV per nucleon. 
The present analysis yields a maximum of 4.5 helium 
nuclei near $Z_{bound}$ = 30 whereas, with the acceptance of the TOF wall
in the previous experiment at 600 MeV per nucleon, 
only about 3.5 helium nuclei were detected
at the maximum \cite{botv2}. 
Lithium is by far the most abundant element in the range of fragments
with intermediate mass. Its maximum multiplicity is 1.6 at
$Z_{bound}$ = 35. With increasing $Z$ the maxima of the 
multiplicity distributions continue to shift
to larger values of $Z_{bound}$. For elements up to $Z$~=~10 
this is shown in fig.~\ref{zdepen}.

\begin{figure}[p]
     \centerline{
     \epsfig{file=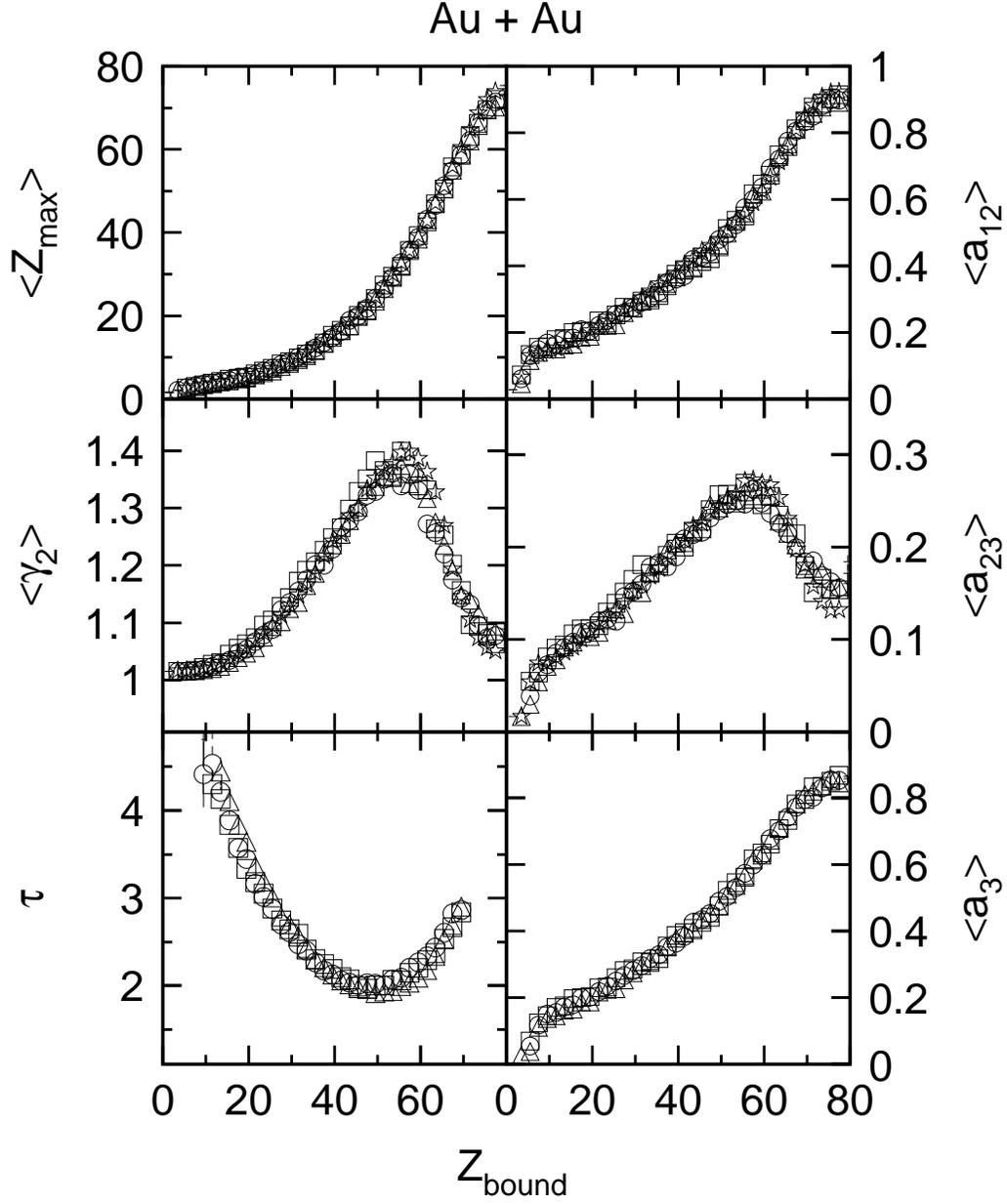,height=17.cm}}
     \begin{center}
     \parbox{15.cm}{\caption[Z-Correlation]{\em
Six charge-correlation 
observables as a function of $Z_{bound}$ for the reaction
$^{197}$Au on $^{197}$Au.
The comparison is made for the four incident energies 
$E/A$ = 400, 600, 800, and 1000 MeV in the top four panels
and for three incident energies $E/A$ = 600, 800, and 1000 MeV
in the two bottom panels.
The definitions are given in
the text, the symbols have the same meaning as in fig. 6.
     \label{zcorrel}
     }}
     \end{center}
\end{figure}

The invariance with respect to the bombarding energy also holds 
for other charge correlations that have been found useful to
characterize the population of the partition space in the fragmentation
process \cite{kreutz}. Six of these 
observables, as measured at the four bombarding
energies, are given in fig.~\ref{zcorrel} as a function of 
$Z_{bound}$. They include the mean values of 
the maximum fragment charge $Z_{max}$ of the event and of three asymmetry
variables. The latter are
the charge asymmetry $a_{12} = (Z_{max}-Z_2)/(Z_{max}+Z_2)$ 
between the two largest fragments, 
the charge asymmetry $a_{23}  = (Z_2-Z_3)/(Z_2+Z_3)$ 
between the second and third largest 
fragment, and the three-fragment charge asymmetry 
\begin{equation}
a_{3} = \frac {\sqrt{ (Z_{max}-\langle Z \rangle)^2
                       +(Z_{2}-\langle Z \rangle)^2
                       +(Z_{3}-\langle Z \rangle)^2}}
               {\sqrt{6} \cdot \langle Z \rangle }
\label{EQ2}
\end{equation}
where
\begin{equation}
\langle Z \rangle = \frac{1}{3} \cdot (Z_{max}+Z_{2}+Z_{3})
\label{EQ3}
\end{equation}
denotes the mean value of the three largest atomic numbers.
The quantity $a_3$ assumes values near one when the partition is
dominated by one heavy residue (large $Z_{max}$) and a
value of zero when the three fragments are of equal size. 
Events containing at least two fragments with $Z \ge$ 2 were selected
for $\langle a_{12} \rangle$, and events containing at least three fragments 
with $Z \ge$ 2 were selected for $\langle a_{23} \rangle$ and 
$\langle a_{3} \rangle$.

The mean value of the normalized width of the fragment-charge
distribution 
\begin{equation}
\gamma_2 = 1 + \sigma^2(Z)/\langle Z \rangle ^2
\label{EQ4}
\end{equation}
peaks at $Z_{bound}$ = 55 which is
slightly larger than reported previously \cite{kreutz} for the reasons
given above (fig.~\ref{zcorrel}, middle, left). 
The $Z$ spectra measured for given bins of $Z_{bound}$ were fitted with
power-law functions $\sigma(Z) \propto Z^{-\tau}$ over the range
of intermediate-mass fragments.
%3 $\le Z \le Z_P/3$ where $Z_P$ is the projectile atomic number. 
The resulting $\tau$ values (fig.~\ref{zcorrel}, bottom, left)
reach a minimum of $\tau$ = 2.0 in the $Z_{bound}$ range of 45 to 50. 
The characteristic
dependence on the centrality of the collision is in good agreement with
the results reported earlier \cite{ogilvie,kreutz}. 

Once more, we emphasize the striking invariance of the partition patterns
with the incident energy as evident from figs.~\ref{uniener} and~\ref{zcorrel}. 
Although not
specifically 
demonstrated in fig.~\ref{zdepen}, it is also valid for the mean multiplicities
of the individual elements.\\

{\bf 3.3 Accuracy of mean fragment multiplicity}
\vspace{0.2cm}

The attainment of a maximum of 
relative accuracy in the measurements at different 
bombarding energies was one of the motivations for 
adopting a common definition of the
spectator source. The absolute accuracy of the measured fragment 
multiplicities and charge correlations is influenced by several 
experimental effects. They will be listed and discussed in 
the following where we will focus on the maximum
value of the $\langle M_{IMF} \rangle$ versus $Z_{bound}$ correlation.

The upstream beam detectors, made of plastic scintillator foils with a
resultant thickness of together
160 $\mu$m, represented a target of about 2$\cdot 10^{-3}$ 
interaction probability (not taking into account the 
hydrogen content).
Coincident interactions of a beam particle in the scintillator foils 
and in the
target are negligible. However, interactions in the scintillators may
potentially lead to a valid trigger, although the trigger probability is
relatively small due to the reduced solid
angle subtended by the hodoscope with respect to the beam detectors.
The recorded fragment multiplicities of these events, as a function of
the recorded $Z_{bound}$, is slightly lower than that of reactions 
in the target, again due to the reduced solid angle.
From analyzing data sets taken without a target,
we find that the contamination caused by fragmentation reactions in the 
scintillator foils is indeed negligible. For the Au and U targets,
since their interaction probabilities are the lowest, 
the relative contamination is the largest. Even in these cases, the 
contamination causes a reduction 
of the maximum mean multiplicity 
of only about 0.05 to 0.2 units,
somewhat depending on the bombarding energy.

Secondary interactions within the target may change the fragment 
multiplicity on the percent level. The probability is largest 
for the lighter
targets because of their larger interaction probability (up to 3.9\%)
but more violent interactions may occur in the heavier targets.
The geometric acceptance of the Si-CsI(Tl) hodoscope relative to the
subtended solid angle was 85\%. The corresponding
loss of light fragments, at the lower bombarding energies, can reach 0.1 
units at the maximum. With the given detector geometry and adopted limit in
rapidity (section 3.1) there was no dead area between
the acceptances of the hodoscope and of the ALADIN magnet with the TOF 
wall.

Interactions with the pressure window (80-mg/cm$^2$ areal density,
corresponding to about 1\% interaction probability for heavy and less
for lighter fragments) 
and with the counting gas behind the 
magnet may lead to additional secondary fragmentations of produced
fragments, thereby causing both gains and losses in the number
of intermediate-mass fragments. Finally, 
the remaining inefficiencies of the algorithms used for the
TOF-wall analysis may
lead to a misidentifications of fragments if secondary interactions
within the wall are not recognized or double hits are not resolved. 
These effects may produce errors of different sign.

%The analysis of this, in total, fairly long list of mostly very small 
%and partially cancelling effects led to the decision 
%that no individual corrections should be applied.
The estimate of the overall systematic uncertainty of the 
maximum mean fragment multiplicity yielded 
$\Delta \langle M_{IMF} \rangle = \pm$ 0.2 (standard deviation).
At the lower bombarding energies, caused by the fragment detection
with the Si-CsI(Tl) hodoscope, the upper limit of the uncertainty
increases up to $\Delta \langle M_{IMF} \rangle$ = +0.4.\\ 

{\bf 3.4 Dependence on projectile and target mass}
\vspace{0.2cm}

\begin{figure}[tb]
     \centerline{
     \epsfig{file=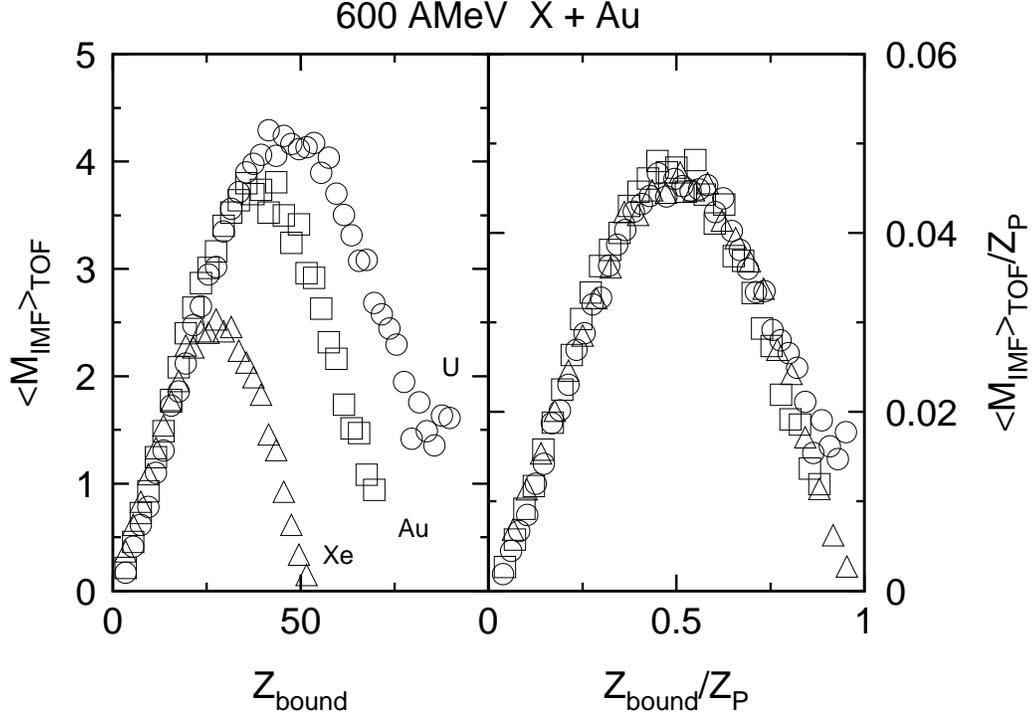,height=10.cm}}
     \begin{center}
     \parbox{15.cm}{\caption[Scaling XeAuU]{\em
Left panel: 
Mean multiplicity of intermediate-mass
fragments $\langle M_{IMF} \rangle_{TOF}$, observed with the TOF wall,
as a function of $Z_{bound}$ for the reactions
$^{238}$U on $^{197}$Au (circles), $^{197}$Au on $^{197}$Au (squares),
and $^{129}$Xe on $^{197}$Au (triangles) at $E/A$ = 600 MeV. Note that
also in $Z_{bound}$ only fragments detected with the TOF wall are included.
\noindent
Right panel:
The same data, as shown in the left panel, after normalizing both 
quantities with respect to the
atomic number $Z_{P}$ of the projectile.
     \label{scaling}
     }}
     \end{center}
\end{figure}

The $\langle M_{IMF} \rangle$ versus $Z_{bound}$ correlation depends 
on the mass of the projectile. The results obtained with the three 
projectiles $^{129}$Xe, $^{197}$Au, and $^{238}$U 
at 600 MeV per nucleon show that,
on the absolute scale, more fragments are produced in the decay of 
heavier projectiles (Fig.~\ref{scaling}, left-hand side). 
However, a normalization with respect to the atomic number 
$Z_P$ of the projectile 
reduces the three curves to a single universal relation 
(Fig.~\ref{scaling}, right-hand side).
Only TOF-wall data
were used for this comparison as the pulse heights from the hodoscope 
were not recorded 
during the runs with the xenon beam.
In addition, the upper limit of the
$Z$ range adopted for intermediate-mass fragments was changed
from $Z \leq$ 30 to $Z \leq Z_{P}/$3.
This is meant to exclude fragments from binary-fission events.
Since the elemental yields
decrease rapidly with $Z$ the exact location of this upper limit 
does not crucially affect this comparison of fragment multiplicities.

\begin{figure}[p]
     \centerline{
     \epsfig{file=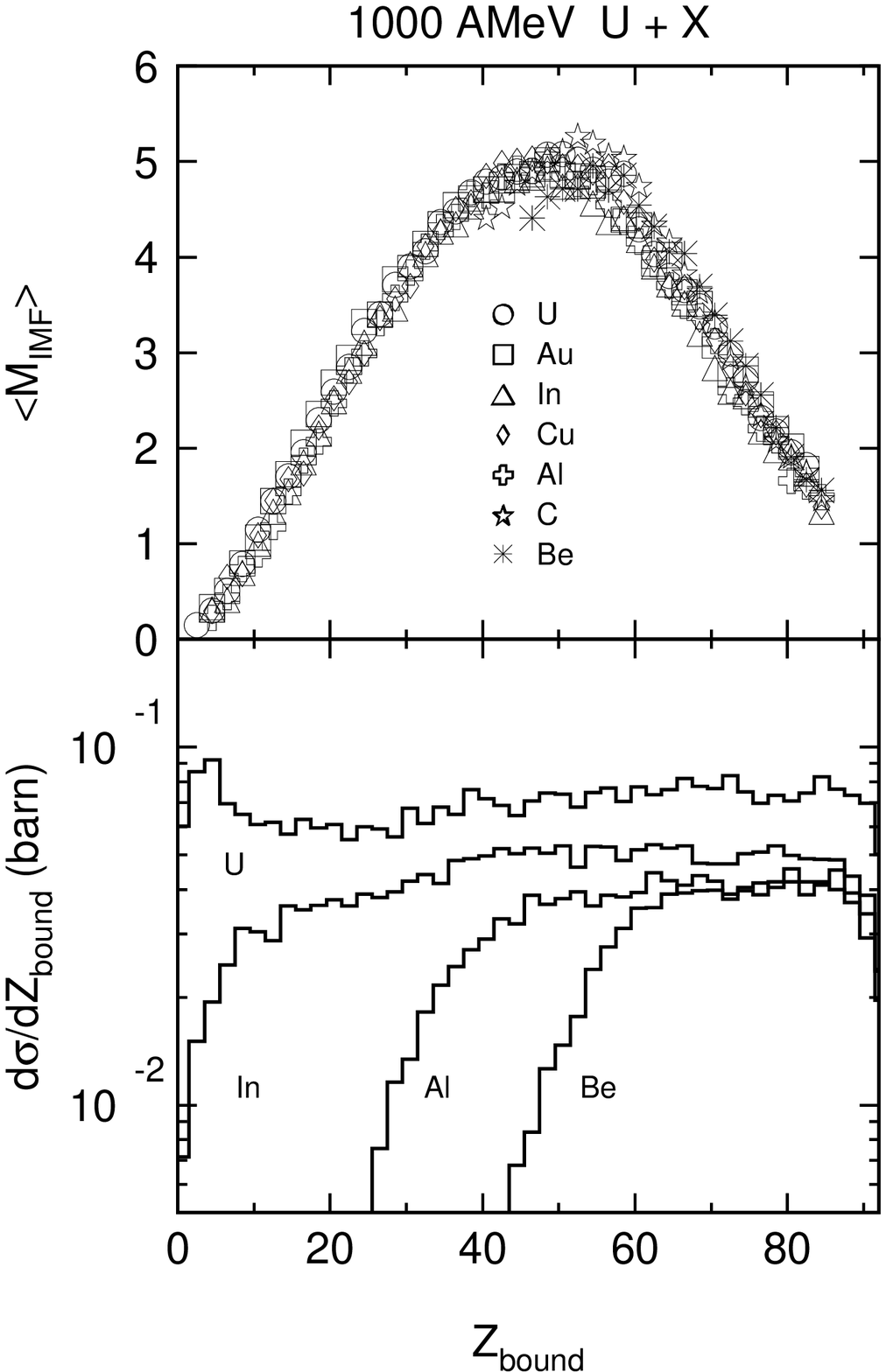,height=16.cm}}
     \begin{center}
     \parbox{15.cm}{\caption[Target Univer U1000]{\em
Top: Mean multiplicity of intermediate-mass
fragments $\langle M_{IMF} \rangle$ as a function of
$Z_{bound}$ for the reactions of $^{238}$U projectiles 
at $E/A$ = 1000 MeV with the seven
targets of Be, C, Al, Cu, In, Au, and U. 
\noindent
Bottom: 
Measured cross sections $d\sigma /dZ_{bound}$
for the reactions of $^{238}$U projectiles 
at $E/A$ = 1000 MeV with the four targets
of Be, Al, In, and U.
Note that the experimental trigger, for the case of uranium beams, 
affected the cross sections
for $Z_{bound} \ge$ 70.
     \label{uniutarg}
     }}
     \end{center}
\end{figure}

The target invariance of the $M_{IMF}$ versus $Z_{bound}$ correlation
was first observed for collisions of $^{197}$Au projectiles 
with C, Al, Cu, and Pb targets at 600 MeV per nucleon [1-4].
%\cite{ogilvie,hubele1,kreutz,hubele2}. 
In fig.~\ref{uniutarg} (top) the universal
nature of this correlation is demonstrated
for $^{238}$U projectiles at 1000 MeV per nucleon 
and for the full set of seven targets. 
With $^{238}$U the maximum mean number of fragments 
(3 $\le Z \le$ 30) is about five
at $Z_{bound}$ = 50. The spread over one half unit
may not be significant in view of the systematic errors discussed
in the last section.
However, a slight tendency towards smaller multiplicities for
the heavier targets can be explained 
by the larger Coulomb deflection that is caused by these targets 
and its effect on the acceptance for the very light fragments \cite{voli}.

\begin{figure}[tb]
     \centerline{
     \epsfig{file=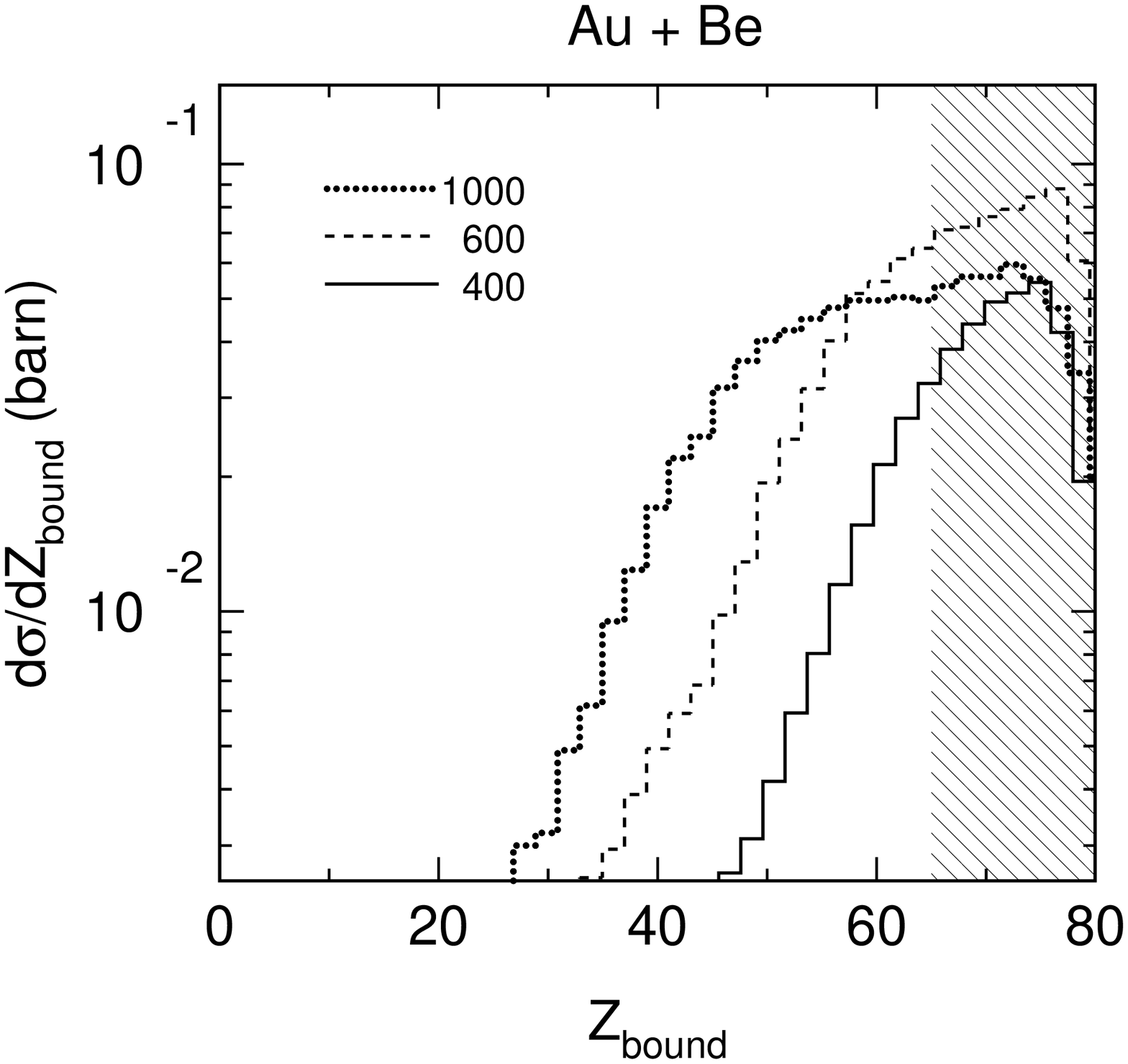,height=9.cm}}
     \begin{center}
     \parbox{15.cm}{\caption[Cross section]{\em
Measured cross sections $d\sigma /dZ_{bound}$
for the reaction $^{197}$Au on Be at $E/A$ = 400 (full line),
600 (dashed), and 1000 MeV (dotted).
Note that the experimental trigger conditions, which were not identical at 
the three bombarding energies, start to affect the cross sections
at $Z_{bound}$ between 60 and 70 (shaded area).
     \label{cross}
     }}
     \end{center}
\end{figure}

The reaction cross sections $d\sigma/dZ_{bound}$ are strongly 
target dependent \cite{hubele1, botv2}. For $^{238}$U projectiles of
1000 MeV per nucleon this is shown in the lower part of fig.~\ref{uniutarg}.
Only part of the $Z_{bound}$ range can be covered with the lighter targets.
The variation of $d\sigma/dZ_{bound}$ with
bombarding energy is insignificant for the collisions with heavy targets.
There, the limit above which the spectator excitation is dominated by the
collision geometry seems to be already reached. With the lighter targets, 
the higher bombarding energies allow for considerably larger energy transfers
in central collisions. For $^{197}$Au on Be, this is demonstrated in
fig.~\ref{cross}. Collisions resulting in $Z_{bound} \approx$ 40 and maximum
fragment multiplicity (cf. fig. 6) are initiated with significant 
probability only at 1000 MeV per nucleon.

\begin{figure}[tb]
     \centerline{
     \epsfig{file=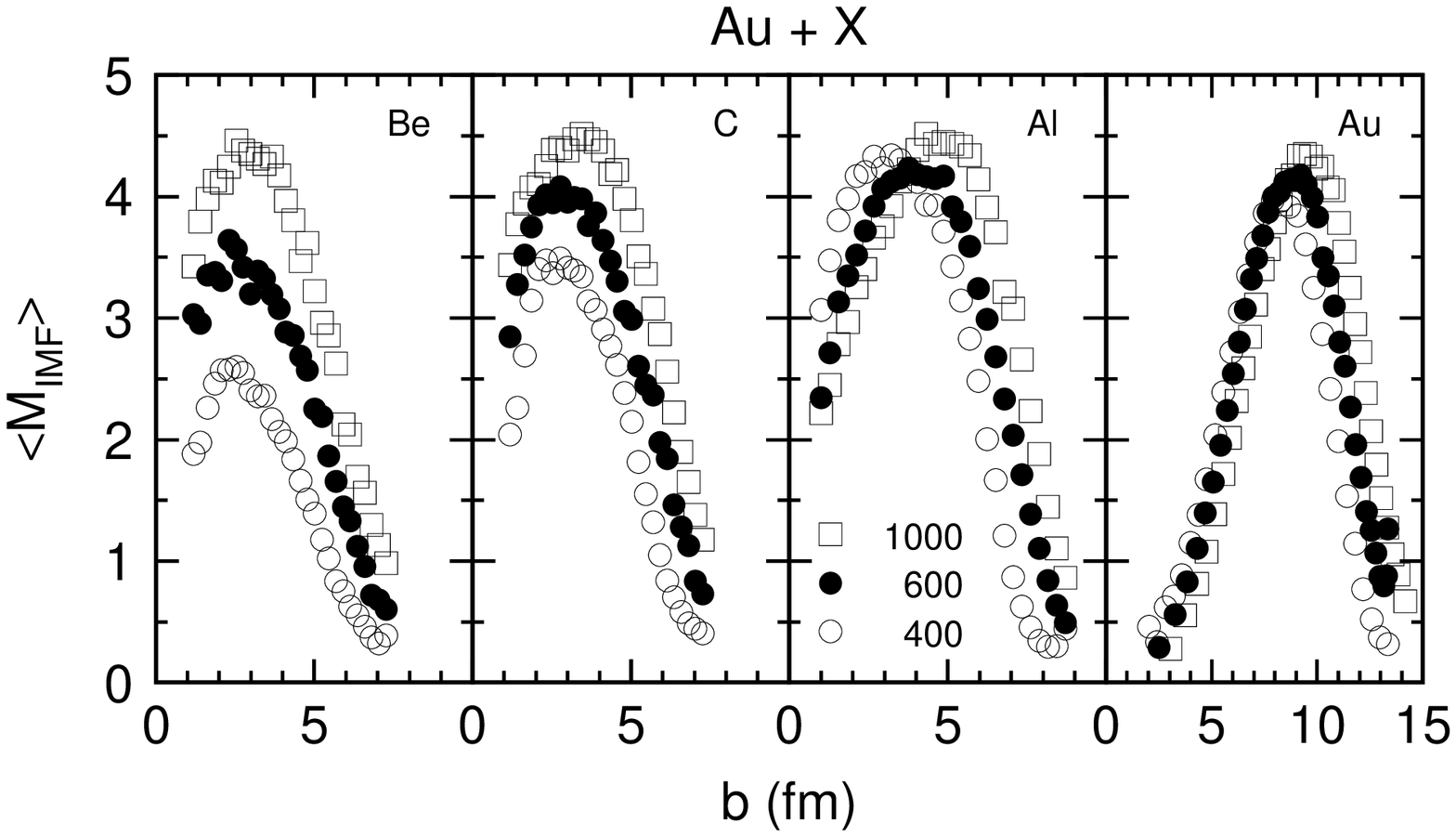,height=9.cm}}
     \begin{center}
     \parbox{15.cm}{\caption[Impact parameter]{\em
Mean multiplicity of intermediate-mass
fragments $\langle M_{IMF} \rangle$ as a function of the
impact parameter $b$, as
deduced from $Z_{bound}$, for the reactions $^{197}$Au on Be, C, Al,
and Au at three energies $E/A$ = 400 (circles), 600 (dots), 
and 1000 MeV (squares). 
     \label{impact}
     }}
     \end{center}
\end{figure}

The same characteristic behavior is illustrated in an alternative way 
in fig.~\ref{impact}. Here
the mean fragment multiplicities are given as a function of the 
impact parameter $b$ for $^{197}$Au projectiles
and four targets at three bombarding energies.
The impact parameter $b$ was
derived by assuming that $b$ is monotonically
correlated with $Z_{bound}$ (see ref. \cite{hubele1} for further details).
With decreasing target mass the maximum fragment multiplicity is 
reached in more central collisions and the dependence on the bombarding
energy increases.
The data for the beryllium target, in particular, show that,
at energies below 1000 MeV per nucleon, the spectators formed even in
the most central collisions are not sufficiently excited to
undergo a complete disassembly. The apparent decrease of 
$\langle M_{IMF} \rangle$ towards
$b$~=~0 for the very light Be and C targets is more likely to be 
caused by autocorrelations than by the mean behavior at small impact 
parameter. The bins of smallest impact parameter correspond to very
small cross sections and are filled with the tails of events with
the smallest 
values of $Z_{bound}$ for which $M_{IMF}$ is restricted by definition.
This was verified by sorting the data measured at 400 MeV per nucleon
according to the light particle multiplicity determined with the
Miniball/wall and Si-CsI(Tl) hodoscope.
In the latter case no comparable decrease is observed at the 
largest particle multiplicities.\\
 
{\bf 3.5 Decay dynamics}
\vspace{0.2cm}

\begin{figure}[tb]
     \centerline{
     \epsfig{file=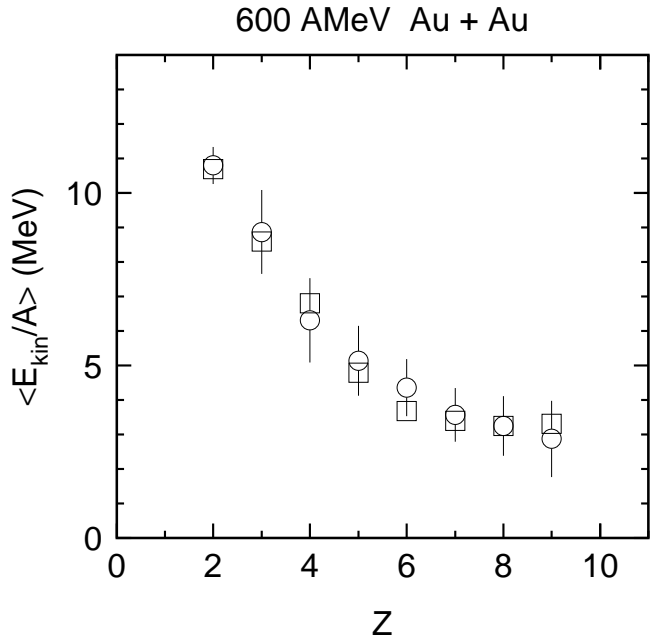,height=9.cm}}
     \begin{center}
     \parbox{15.cm}{\caption[Mean Kin. Energy]{\em
Mean kinetic energies per nucleon in the moving frame, 
deduced from the transverse (circles) and longitudinal (squares)
momentum widths as 
described in the text, for fragments from
the reaction $^{197}$Au on $^{197}$Au at $E/A$ = 600 MeV
and for 20 $\le Z_{bound} \le$ 60.
     \label{meanekin}
     }}
     \end{center}
\end{figure}

The observed invariance of the transverse-momentum widths with 
respect to the bombarding energy (fig.~\ref{transm}) 
indicates that the entrance-channel dynamics play a minor role for the
decay of the excited spectator. An estimate of the mean kinetic energies
of the produced fragments in the moving frame was obtained in the following
way: Event-by-event the vertical positions $r_{yi}$ of all fragments 
measured with the
TOF wall were converted into transverse momenta $p_{yi}$ according to
\begin{equation}
p_{yi} = (r_{yi}/L_i) \cdot A_i \cdot p_{P}/A_{P}
\label{EQ5}
\end{equation}
Here the notation is the same as in eq. 1 with the additional subscript i 
denoting the fragment number within an event.
With this information the center-of-mass motion of the
decaying spectator in y direction 
and the intrinsic velocities of the fragments in this frame,
in the y direction,
were calculated. It was then assumed that the corresponding kinetic 
energies for one transverse
degree of freedom represent one third of the total kinetic
energies in the moving frame. 
Results for the reaction $^{197}$Au on
$^{197}$Au at 600 MeV per nucleon, integrated over 
20 $\le Z_{bound} \le$ 60, are given in fig.~\ref{meanekin} (open circles).
Because of the absence of a noticeable dependence on the 
bombarding energy (cf. fig.~\ref{transm}) they
are representative for the whole energy range over which the universal
spectator decay prevails.
The assumption of dynamical equilibration over the three degrees of
freedom was verified by using the measured velocities 
for the corresponding analysis in the longitudinal direction. 
The result, given by the open squares in fig.~\ref{meanekin},
turned out to be identical, within the errors.

The mean kinetic energies per unit fragment mass 
$\langle E_{kin}/A \rangle$ decrease 
rapidly with atomic
number $Z$.
In the limit of purely thermal contributions to the
kinetic energies, $\langle E_{kin}/A \rangle$ is expected to have 
a 1/$A$ dependence
which is approximately observed. However, on the order of one half
of the kinetic energies in the rest frame of the decaying system may 
originate from Coulomb repulsion and sequential decays of excited fragments
\cite{voli}. With this assumption the magnitude of the kinetic
temperature $T = 2/3 \cdot 1/2 \cdot \langle E_{kin} \rangle$ assumes
a value of approximately 15 MeV. This exceeds considerably 
the emission temperatures $T \approx$ 5 MeV derived from the relative 
isotopic abundances \cite{pocho1} or from relative yields of particle
unbound states \cite{kunde1} which represents a well known but up to now not 
fully resolved problem [28-31].
%\cite{barz2,boal,barz3,bauer1}.
Quantitatively, these values of the kinetic and the emission temperatures
are in good agreement with those calculated by Bauer \cite{bauer1} 
who assumes that the higher kinetic temperatures reflect 
the additional Fermi momenta of the constituent nucleons 
of a fragment \cite{goldhaber}. 
Small collective contributions to the kinetic energies can also 
have large effects on the apparent temperatures
\cite{milkau,botv3}.
On the other hand, the comparison with central collisions 
of $^{197}$Au on $^{197}$Au at 100 MeV per nucleon, 
made in ref. \cite{kunde2},
shows that, in the present case of spectator decay, 
the upper limit for possible 
dynamical contributions is rather small.
The kinetic energies are thus predominantly indicative of a 
thermally driven breakup out of an expanded state.

\section{Discussion}
\label{Sec_4}

{\bf 4.1 Fragmentation at high bombarding energies}
\vspace{0.2cm}

Exclusive fragmentation studies at high bombarding energies were
performed by several groups [36-43].
%\cite{wadding,lips,kwiat2,rusch,jain,jain2,klmm1,klmm2}.
The experiments using light projectiles of mass $A \le$ 4
confirm the finding that very high bombarding energies are needed
in asymmetric systems in order to achieve the complete disassembly of a
heavy spectator nucleus with large probability \cite{lips,kwiat2}.
These data fit rather well into the observed variation of the 
maximum mean multiplicity in central collisions with light 
targets (section 3.4). 

\begin{figure}[tb]
     \centerline{
     \epsfig{file=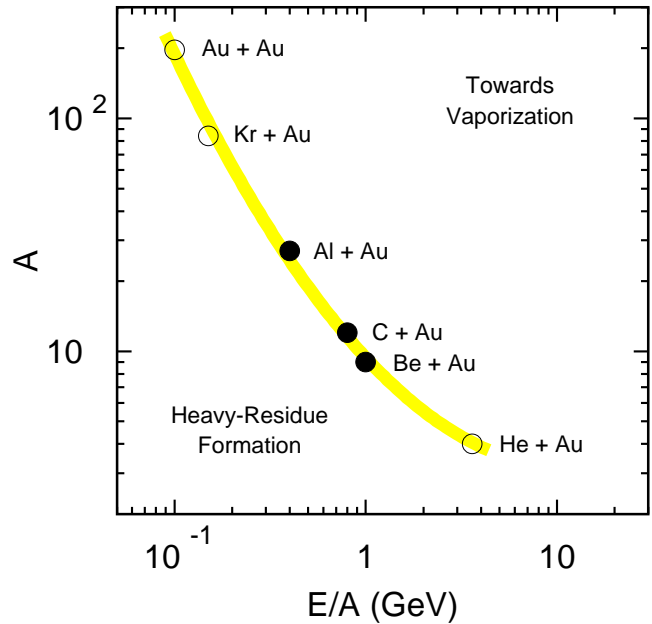,height=9.cm}}
     \begin{center}
     \parbox{15.cm}{\caption[Limiting Fragmentation]{\em
Mass number $A$ of the collision partner versus bombarding energy
for reactions of $^{197}$Au nuclei for which maximum fragment production has 
been observed in central collisions.
Full points denote reactions
studied in this work in reverse kinematics, 
open symbols refer to work reported in
\cite{tsang,lips,peaslee}.
     \label{limfrag}
     }}
     \end{center}
\end{figure}

A schematic representation of this systematic behavior, for reactions 
of gold nuclei with partners of different mass, is given in fig.~\ref{limfrag}.
It includes the results obtained for $^{197}$Au on $^{197}$Au 
at lower energies \cite{tsang} and for $^{84}$Kr on $^{197}$Au \cite{peaslee}.
The figure shows the relation between the bombarding energy that is needed
in order to observe maximum fragment production in
central collisions and the mass number 
of the collision partner. For the reactions studied in this work in reverse
kinematics, these energies may be taken from fig.~\ref{impact}: 
for $^{197}$Au on C, e.g., the maximum fragment multiplicity is 
not reached with 600 
but is easily reached with 1000 MeV per nucleon incident energy. In fact, 
already at 800 MeV per nucleon a considerable cross section is associated
with a mean fragment multiplicity of $\approx$ 4.5 
(data not shown in fig.~\ref{impact}). 

The threshold energies of the other reactions
were determined in a similar way. 
In the case of the $^{4}$He on $^{197}$Au 
reactions, a moderately complete excitation function is not available
and the given value of 3.6 GeV per nucleon may represent an upper limit.
At this energy the maximum of fragment production is reached,
as has been argued on the basis of both the fragment multiplicity
and the observation of a minimum in the $\tau$ parameter describing the 
elemental yields \cite{lips}. Light-particle induced reactions at 
significantly lower
energies are incapable of producing spectators at sufficiently high
excitation. This is evident from the excitation energies of $E_x/A \leq$
3 MeV for collisions of $^1$H and $^3$He projectiles of 2-GeV total
energy with gold targets, 
deduced from neutron multiplicity measurements \cite{pienko},
and consistent with the relatively small mean fragment multiplicities 
reported for
$^3$He on $^{197}$Au at 1.6 GeV per nucleon \cite{kwiat2} and 
$^4$He on $^{197}$Au at 1.0 GeV per nucleon
\cite{lips}. For this latter group of reactions, the fragment
multiplicity is still on the rising side where the partitions are
characterized by one large fragment (large $Z_{max}$) and a correspondingly
large first asymmetry $a_{12}$ (cf. fig.~\ref{zcorrel}). This is the regime of
residue formation (fig.~\ref{limfrag}). On the other hand, with collision 
partners of masses or energies exceeding the optimum values, the central
collisions will lead to spectators of decreasing mass and to their
disassembly into an increasingly larger number of
light fragments and particles and, eventually, to vaporization
\cite{tsang,reisdorf}. 

\begin{figure}[tb]
     \centerline{
     \epsfig{file=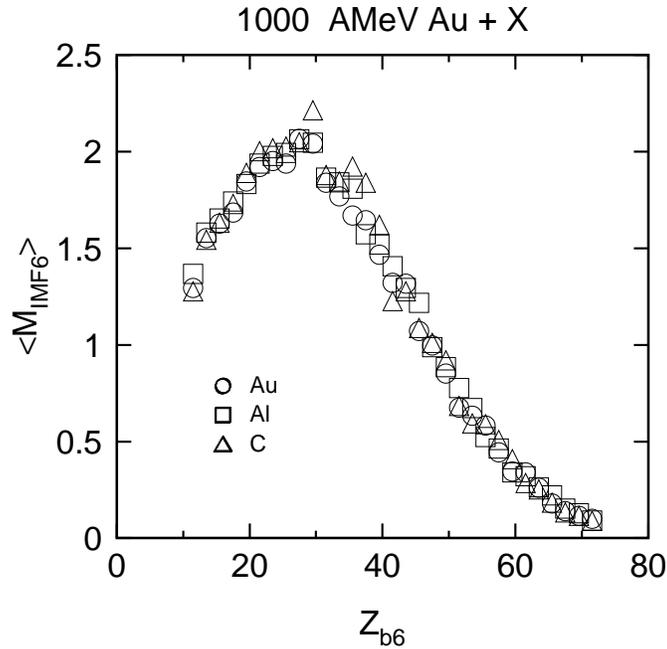,height=9.cm}}
     \begin{center}
     \parbox{15.cm}{\caption[IMf vs Zb6]{\em
Mean multiplicity $\langle M_{IMF6} \rangle$ of intermediate-mass
fragments with $Z \ge$ 6
as a function of the bound charge $Z_{b6}$ (see text) 
for the reactions $^{197}$Au on C, Al, 
and Au at $E/A$ = 1000 MeV.
     \label{zb6}
     }}
     \end{center}
\end{figure}

High charge resolution and full phase space coverage above the
identification threshold of $Z \ge$ 6 has been
achieved with plastic nuclear track detectors \cite{rusch,dreute}.
For a comparison with the results of the Siegen group \cite{rusch},
obtained at the Bevalac with $^{197}$Au beams in the range of 0.85
to 1 GeV per nucleon,
a threshold of $Z \ge$ 6 was applied to our data. The sorting variable
$Z_{bound}$ was replaced by $Z_{b6}$ which represents the sum of the
atomic numbers $Z_i$ of all fragments with $Z_i \ge$ 6.
The correlation between the mean multiplicities $\langle M_{IMF6} \rangle$ 
of fragments with 6 $\le Z \le$ 30 and $Z_{b6}$ is shown in fig.~\ref{zb6}.
A maximum of $\langle M_{IMF6} \rangle \approx$ 2 
is reached for $Z_{b6}$ between 20 and 30.
The data measured with carbon targets in the two experiments agree 
quantitatively with each other within the quoted errors (cf. fig.~\ref{zdepen}
of ref. \cite{rusch}). This mutually confirms the experimental and
analysis techniques. The small reduction of $\langle M_{IMF6} \rangle$
with increasing mass of the target, reported by the Siegen group, is
less pronounced in our data (figs.~\ref{uniutarg} and~\ref{zb6}). 
The mean multiplicities
$\langle M_{IMF6} \rangle$ in the range 20 $\le Z_{b6} \le$ 30
differ by 4\% $\pm$ 2\% for the C and Au targets. In fact, 
we find that the 
invariance with respect to the target and with respect to 
the bombarding energy holds
for arbitrary choices of a lower threshold in $Z$.
This is true also for the correlation observables displayed in
fig.~\ref{zcorrel}.

Emulsion targets were used by Jain and Singh in their study of the
fragmentation of $^{84}$Kr beams of 1.52 GeV per nucleon, 
performed at the Bevalac \cite{jain}. They observe a maximum
multiplicity of $\langle M_{IMF} \rangle$ of 1.8 to 2.0. This
corresponds to $\langle M_{IMF} \rangle / Z_{P}$ = 0.053 which is
in excellent agreement with the present results for Au and U 
beams (figs. 6 and 10), thus extending the invariance with
respect to the bombarding energy and the projectile mass over
a wider range. Note that the slightly lower multiplicities 
shown in fig.~\ref{scaling},
measured with the TOF wall at 600 MeV per nucleon incident energy,
do not include the fragments detected with the hodoscope.

In their work, performed at the AGS in Brookhaven, Jain {\it et al.} 
have studied the fragmentation 
of $^{197}$Au projectiles at 10.6 GeV per nucleon \cite{jain2}.
The fragment charges were deduced from $\delta$-ray counts. The fragment
multiplicities, charge asymmetries, and widths of the charge distributions 
at this high energy are very similar to those measured 
at 1 GeV per nucleon and below. Jain {\it et al.} show a comparison with 
the results reported by Hubele {\it et al.} \cite{hubele1}, 
measured at 600 MeV per nucleon and with the 
acceptance given by the TOF wall of 1-m length.
With the acceptance offered by the extended TOF wall
and with the source definition adopted in the present 
work, the maximum of the $\langle M_{IMF} \rangle$ versus $Z_{bound}$ 
correlation is at $Z_{bound}$ = 40 which coincides with the result 
found at 10.6 GeV per nucleon. 
However, the maximum number of intermediate-mass fragments of 3.5, 
as reported by Jain {\it et al.}, 
is about one unit below the values obtained in
the present work. 

The fragmentation of $^{197}$Au projectiles of 10.6 GeV per nucleon
in collisions with emulsion targets
was also investigated by the KLMM Collaboration \cite{klmm1,klmm2}.
In their more recent report \cite{klmm2} 
these authors present a rather detailed 
comparison with the data of
Hubele {\it et al.} \cite{hubele1} and Kreutz {\it et al.} \cite{kreutz}.
Some of the conclusions are modified if the comparison is 
made with the new data of this work.   
We find that the correlations of $Z_{max}$ with $Z_{bound}$,
measured at 10.6 GeV per nucleon and at $\le$ 1 GeV per nucleon,
agree now perfectly. Good agreement is also observed for the 
charge variance $\langle \gamma_2 \rangle$ (eq. 4). 
However, the maximum value of 
$\langle M_{IMF} \rangle$ of 3.2, reached at $Z_{bound} \approx$ 45 
and in good agreement with the data of ref.
\cite{jain2}, is one unit lower than measured in this work.
By comparing the multiplicities as a function of $Z$ we have identified
the difference as mainly occuring in the yields of very light fragments.
A maximum mean multiplicity for 3 $\le Z \le$ 6 of 1.8 is reported
for 10.6 GeV per nucleon while it is
3.0 in our data in the same $Z$ range 
(cf. fig.~\ref{zdepen}). The difference in the maximum mean He
multiplicity (6.5 at the higher and 4.5 at the lower bombarding
energies) may be
partly due to the rapidity limit applied in the present work 
(fig.~\ref{rapid}).
It is not excluded, at present, that the differences may reflect a weak 
variation of the fragmentation pattern with
bombarding energy; on the other hand,
emulsion data measured at 1 GeV per nucleon \cite{wadding}
were shown to agree with those for 10.6 GeV per nucleon 
over a major part of the $Z_{bound}$ range \cite{klmm2}.

The KLMM collaboration has reported 
$Z_{bound}$ distributions which are sorted individually
for reactions on the light (H, C, N, and O) and on the heavy constituents 
of the emulsion target \cite{klmm1}. Qualitatively, the
distribution for the light constituents decreases
with decreasing $Z_{bound}$ in a similar way as the distributions for light
targets in the present work. On the more quantitative level,
however, the closest resemblance is with the distribution measured 
for the $^{197}$Au on Al reaction at
1000 MeV per nucleon rather than the $^{197}$Au on C reaction at this energy. 
This might be considered as another indication of the role of the 
bombarding energy in reaching large cross sections for high-multiplicity 
breakups in asymmetric systems.

Towards higher bombarding energies, no limit for the universality 
of the spectator fragmentation is known at present.
Limiting fragmentation, i.e. a saturation of the
production cross sections for intermediate-mass fragments at bombarding
energies in excess of several GeV, has been observed [47-49].
%\cite{limit1,limit2,porile}. 
It is also an experimentally
established fact that the elemental 
distributions change very little at high bombarding 
energies \cite{berthier,traut}.
This suggests that the cross sections for the formation of highly
excited spectator nuclei may be rather weak functions of the
bombarding energy in the GeV range. In a pure abrasion picture this is
expected since the cross section and the excitation energy are mutually
interrelated via their geometric dependence on the impact 
parameter \cite{gaimard}.
On the other hand, it is known that the geometric abrasion model can
only account for part of the excitation energy deposited in the formed
spectator nuclei \cite{brohm}. The missing part is thought to be due to
struck nucleons and secondary particles being scattered into the 
cold spectator matter. Some of these particles 
may be deltas and other nucleon resonances which carry additional
energy. It is known from experiment that the cross sections and transverse
momentum distributions of scattered or produced particles change very
slowly at high bombarding energies \cite{kno,ziping}. Thus, also these
factors will not introduce a dramatic energy dependence.
One may therefore conclude that the spectator formation and decay, 
as observed in the
present work, persists up to the highest
bombarding energies with virtually invariant features.\\

{\bf 4.2 Model calculations}
\vspace{0.2cm}

Model calculations were performed in order to identify possible reasons
for the observed $Z_{bound}$ universality. The excitation of the primary
spectator nucleus as a function of the number of abraded nucleons was
studied with the intranuclear-cascade model. A recent version of the
ISABEL code by Yariv and Fraenkel with the slow rearrangement option
was used \cite{suemmerer,yariv}. Less emphasis was given to the
absolute values of the excitation energies, which may be overestimated 
by this model (see next section), than to their systematic behavior.

\begin{figure}[p]
     \centerline{
     \epsfig{file=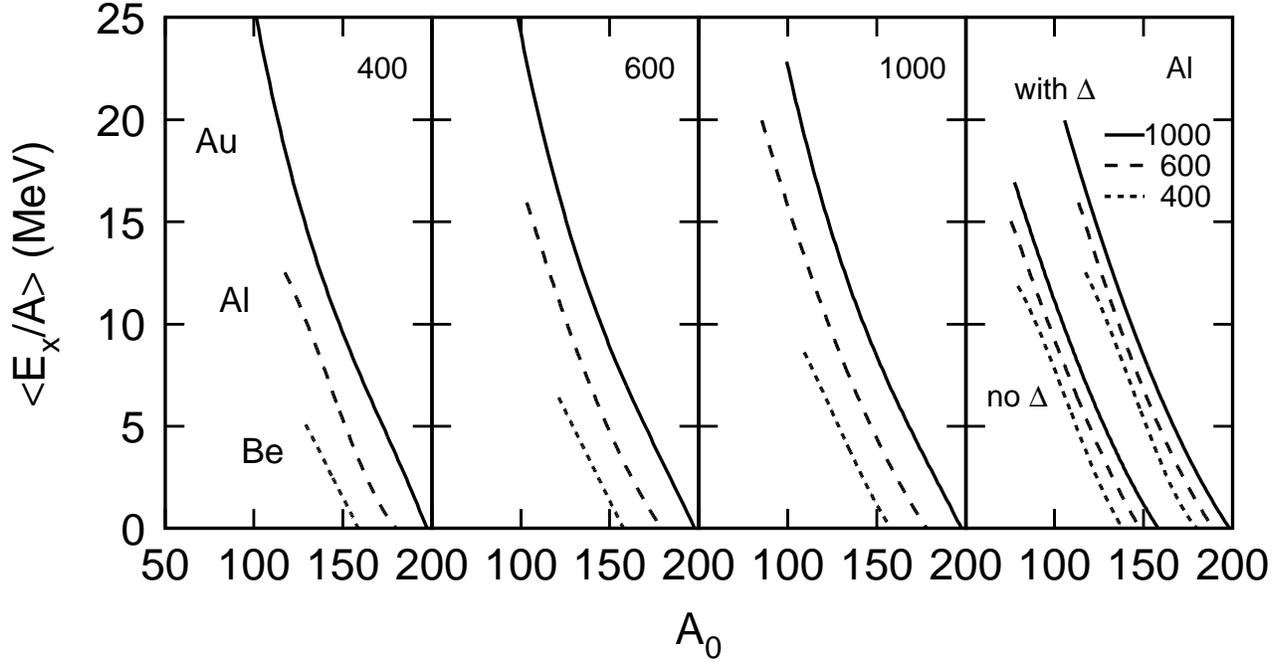,width=17.cm}}
     \begin{center}
     \parbox{15.cm}{\caption[Intra Nuc. Cascade]{\em
Results of the intranuclear-cascade calculations: Mean 
specific excitation energy $\langle E_{x}/A \rangle$ as a function of
the mass number $A_0$ of the primary spectator for the reactions 
$^{197}$Au on Be, Al, and Au targets at three bombarding energies
$E/A$ = 400 MeV (left-most panel), 600 MeV (middle-left), and 1000 MeV
(middle-right). The right-most panel shows the results with and without
$\Delta$ production for the case of the Al target and for
the three incident energies. All curves start at $A_0$ = 197 and have
nearly the same slopes. For the purpose of illustration the following
offsets were applied: in the first three panels 
20 units of $A_0$ for each consecutive target 
and, in the last panel, 10 units of $A_0$
between consecutive lines within a group and 40 units between the two
groups.
     \label{cascade}
     }}
     \end{center}
\end{figure}

\begin{figure}[tb]
     \centerline{
     \epsfig{file=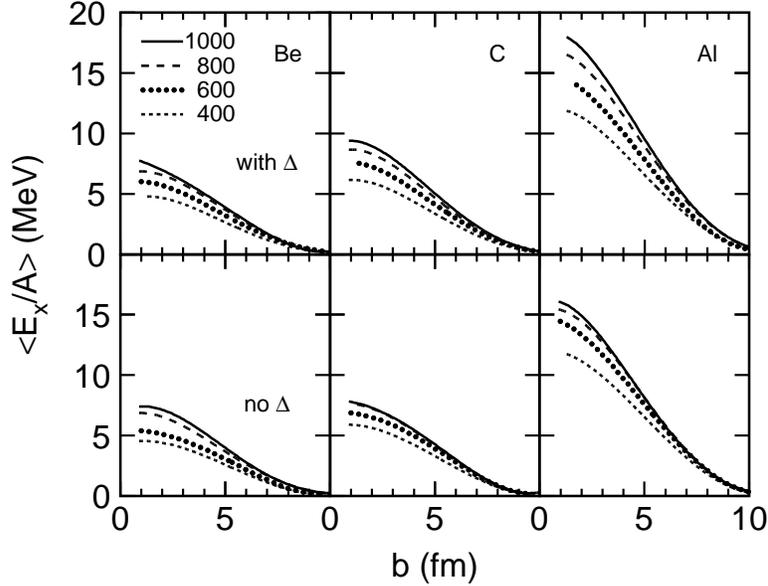,height=9.cm}}
     \begin{center}
     \parbox{15.cm}{\caption[INC Exitation Energy]{\em
Results of the intranuclear-cascade calculations:
Mean specific excitation energy 
$\langle E_{x}/A \rangle$ of the primary spectator
as a function of the impact parameter $b$ for the reactions
$^{197}$Au on Be, C, and Al targets at four bombarding energies
$E/A$ = 400, 600, 800, and 1000 MeV. The upper (lower) row of panels 
shows the results with (without) $\Delta$ production.
     \label{incexi}
     }}
     \end{center}
\end{figure}

\begin{figure}[tb]
     \centerline{
     \epsfig{file=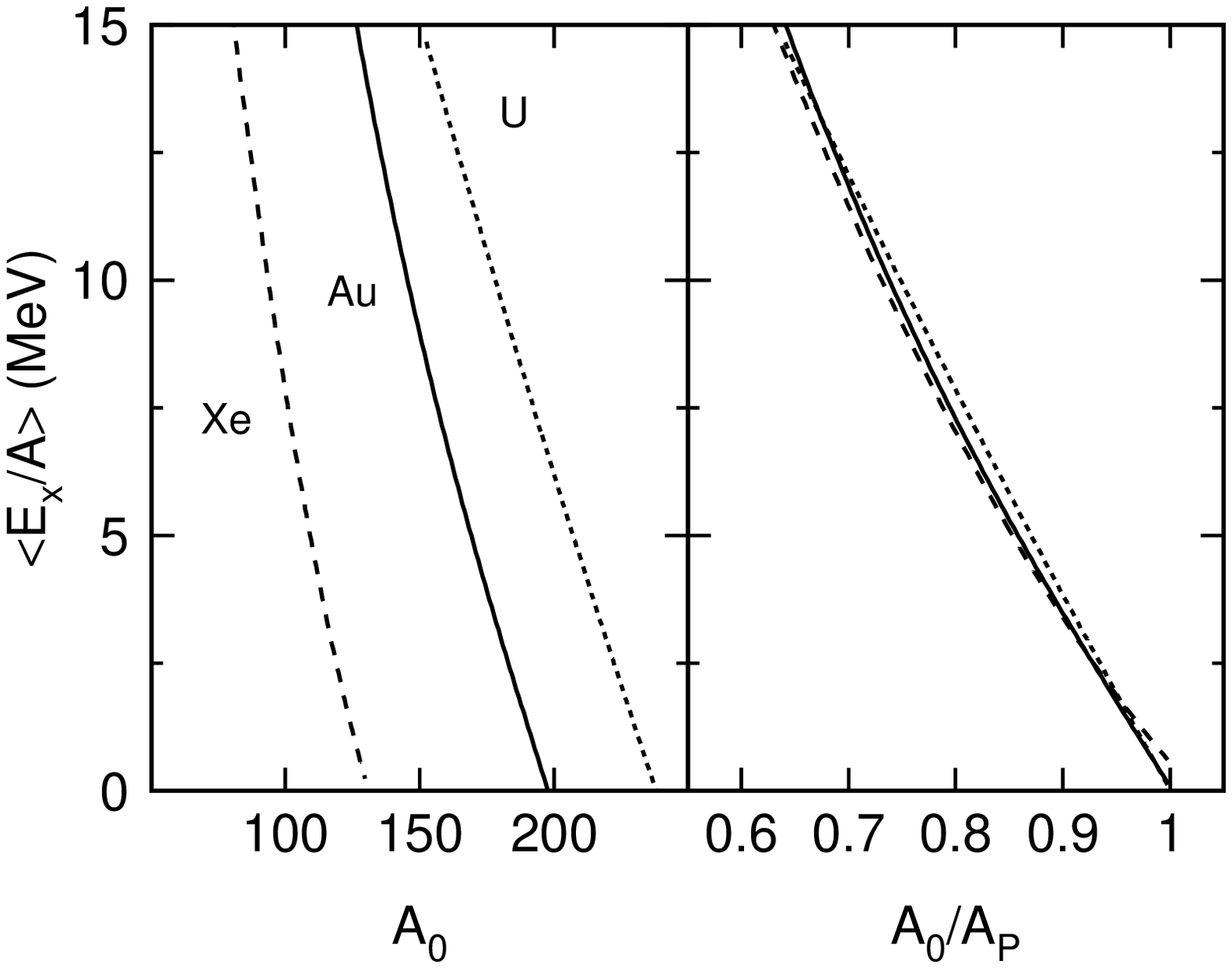,height=9.cm}}
     \begin{center}
     \parbox{15.cm}{\caption[INC Exi vs A0]{\em
Results of intranuclear-cascade calculations:\\
Left panel: Mean specific excitation energy 
$\langle E_{x}/A \rangle$
as a function of the mass $A_0$ of the primary spectator for the reactions of
$^{238}$U, $^{197}$Au, and $^{129}$Xe on a Au target at 
$E/A$ = 600 MeV. The calculations were restricted to impact 
parameters $b \ge$ 8 fm.\\
Right panel: The same correlations plotted as a function of the
reduced mass number $A_0/A_{P}$.
     \label{inca0}
     }}
     \end{center}
\end{figure}

The obtained results, calculated for collisions of $^{197}$Au projectiles
with different targets, reflect the universality seen in the experiment.
The slopes of the correlation of the mean specific excitation energy
$\langle E_x/A \rangle$ with 
the mass $A_0$ of the primary projectile spectator are found
to be independent of the bombarding energy and of the mass of the
target (fig.~\ref{cascade}, note the offsets in $A_0$). Primary spectators of
a given mass seem to always have the same invariant excitation
energy. Therefore, if the subsequent decay proceeds statistically, 
the same universal fragmentation patterns will be observed in the
corresponding regions of $Z_{bound}$.
It is further found that 
the maximum of the specific excitation energy that can be 
reached with a given target depends strongly on the target mass and,
less dramatically, on the bombarding energy (figs.~\ref{cascade}
and~\ref{incexi}). 
Also these features are consistent with the
experimental observations (section 3.4).

\begin{figure}[tb]
     \centerline{
     \epsfig{file=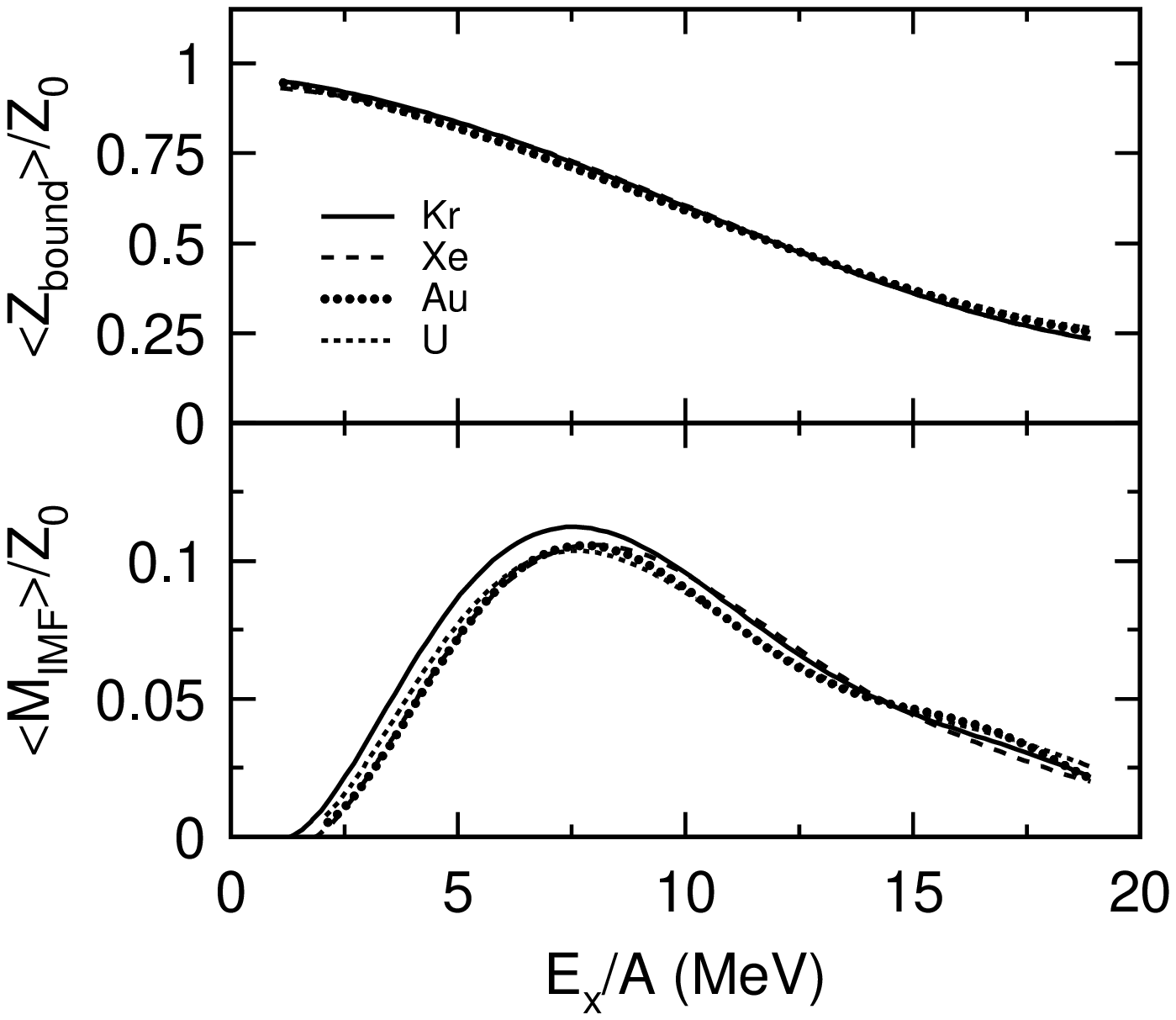,height=9.cm}}
     \begin{center}
     \parbox{15.cm}{\caption[McFRAG MImf vs Exi]{\em
Results of calculations with the statistical multifragmentation model,
performed with the code McFRAG \cite{deangelis}:
Reduced variables $\langle Z_{bound} \rangle /Z_0$ (top) and
$\langle M_{IMF} \rangle /Z_0$ (bottom) as a function of 
the specific excitation energy 
$E_{x}/A$ for the four cases of excited Kr, Xe, Au, and U nuclei.
     \label{mcfrag}
     }}
     \end{center}
\end{figure}

The fairly weak variation with the bombarding energy may seem
somewhat surprising since the excitation of $\Delta$ resonances 
has been suggested to be an efficient means of energy dissipation [57-59].
%\cite{cugnon,brown,kwiat1}. 
The rate of $\Delta$ production grows rapidly with bombarding energy
in the range 400 to 1000 MeV per nucleon, 
as evident from measured pion yields
\cite{harris,muentz}. Consequently, the cross sections for the production
of spectators with high excitation energy may be expected to increase as 
well. This is only partly confirmed by the calculations.
If the code is modified, so as to suppress the production of $\Delta$ 
resonances, the specific excitation energies as a function of the
impact parameter
are only slightly lower, even at the higher bombarding energies 
(fig.~\ref{incexi}). 
There, however, and, in particular for the lightest
targets, the heating by $\Delta$ excitation may be important on a
quantitative level and may determine whether the maximum of fragment
production can be observed in a given collision system (cf. previous 
section). We notice that, in contrast to $\langle E_x/A \rangle$ 
versus $b$, 
the relation $\langle E_x/A \rangle$ 
versus $A_0$ does not depend on whether $\Delta$ production 
is included or suppressed in the calculations (fig.~\ref{cascade}, 
right-most panel). 
It does not respond to this modification of the heating mechanism.
As outlined above, it is essentially the invariance of this function which  
is responsible for the observed $Z_{bound}$ universality. For the case of
hadron induced reactions, this invariant feature has been known since 
the early intranuclear-cascade models were developed \cite{toneev}.

The results of the calculations also exhibit 
the linear scaling with the projectile mass
seen in the experiment (section 3.4). The slopes of 
$\langle E_x/A \rangle$ 
versus $A_0$ grow steeper with decreasing mass of the projectile but are
the same when plotted versus $A_0/A_{P}$ (fig.~\ref{inca0}). This behavior is
expected for the purely geometric abrasion. The 
intranuclear-cascade calculations demonstrate that this feature is not
altered by the additional mechanisms of spectator excitation.

It is not a priori obvious why the linear scaling with the projectile mass
should be preserved during the statistical decay.
The surface energies play an important role in the 
multi-fragment decays, and the ratio of surface to volume of the 
decaying system may not be unimportant. Calculations within the statistical
multifragmentation model show, however, that effects of this origin are not 
significant in the range of projectile masses from krypton to uranium
(fig.~\ref{mcfrag}). The Berlin model in the version implemented in the code 
McFRAG \cite{deangelis} was used. In these calculations the mass $A_0$
and charge $Z_0$ were
kept fixed while $E_x/A$ was varied. This ignores the 
correlation between excitation energy and mass of the primary spectator,
caused by the primary reaction stage, and results in much larger peak
multiplicities than experimentally observed.
The figure, nevertheless, demonstrates that both scaled functions
$\langle Z_{bound} \rangle/Z_0$ and $\langle M_{IMF} \rangle/Z_0$ follow 
universal curves as a function of the specific excitation energy $E_x/A$.
Combined with the intranuclear-cascade relation between 
$\langle E_x/A \rangle$ and $A_0/A_{P}$ (fig.~\ref{inca0}), the model result is
compatible with the observed dependence on the projectile mass.
The top part of fig.~\ref{mcfrag}, furthermore, illustrates how $Z_{bound}$,
according
to the model, increasingly deviates from the charge $Z_0$ of the produced
spectator system as its excitation energy increases.

The link between the first and the later reaction stages is a subject
of high current activity (see, e.g., \cite{botv1,friedman,papp}).
This concerns the expansion and equilibration 
which are thought to have occurred prior to the fragment decay described by 
the statistical multifragmentation models. Intuitively, one may suspect 
that the random knockout of nucleons from and the injection of nucleons 
into the primary spectator produce a highly disordered system capable of
rapidly evolving towards 
equilibrium. The origin of the $Z_{bound}$ universality may thus be 
primarily seen in the stochastic nature of the initial cascade process.\\

{\bf 4.3 Energy deposition}
\vspace{0.2cm}

\begin{figure}[tb]
     \centerline{
     \epsfig{file=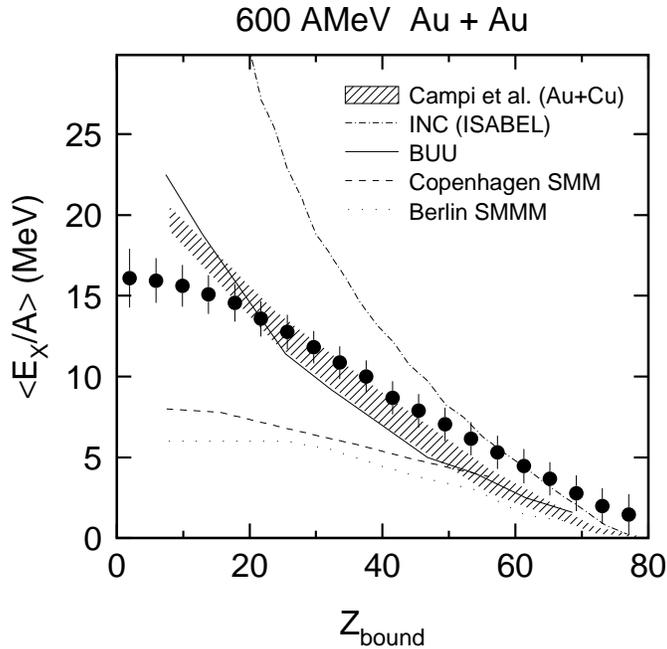,height=9.cm}}
     \begin{center}
     \parbox{15.cm}{\caption[Exi vs Zbound]{\em
Excitation energy per nucleon $E_x/A$ as a function of $Z_{bound}$.
The data points give the results deduced from summing up the breakup
Q values, calculated for the observed partitions,
and the kinetic energies 
for the reaction
$^{197}$Au on $^{197}$Au at $E/A$ = 600 MeV.
The shaded area and the lines represent the results of the analysis by
Campi {\it et al.} \cite{campi}, 
of the intranuclear-cascade 
(this work) and the BUU calculations \cite{kreutz},
and of the analyses with the Copenhagen \cite{barz1} and
Berlin \cite{baoanli} statistical
multifragmentation models, all for the reaction
$^{197}$Au on Cu at $E/A$ = 600 MeV. When necessary, the impact
parameter $b$ was converted into $Z_{bound}$ by using the empirical
relation obtained for the $^{197}$Au on Cu reaction.
     \label{exiwoe}
     }}
     \end{center}
\end{figure}

A quantitative knowledge of the energy transfer to the primary
spectator is indispensable for any interpretation of the multi-fragment
decay in terms of nuclear-matter properties. On the other hand, the 
transient nature of the excitation and decay processes 
makes it difficult to arrive at a consistent definition of the spectator
system and its associated excitation energy. 
The present situation is summarized in fig.~\ref{exiwoe}
which shows the results of 
six analyses for spectators from $^{197}$Au
collisions with either Au or Cu targets as a function of $Z_{bound}$.
The types of analysis are qualitatively different, relying either on
model calculations, on the experimental results or on both.
To permit the comparison in a single figure, 
results calculated as a function of the impact parameter $b$ were converted
to the $Z_{bound}$ scale by using the derived 
empirical relations between the two quantities (section 3.4).

The highest energy deposits in the spectator system are obtained with
the intranuclear-cascade model (cf. previous section). They represent
the sum of the hole energies left behind by nucleons
knocked-out from the spectator
and of the energies carried by struck nucleons captured into the
spectator (the threshold parameter was set to 14 MeV).
Since the evolution of the spectator is ignored
these values may be considered as upper limits.
Considerably lower excitation
energies were obtained with
the Boltzmann-Uehling-Uhlenbeck (BUU, \cite{bauer2}) 
model as reported in ref. \cite{kreutz}. There, the spectator
was defined and analyzed after the system had evolved for a time of
60 to 100 fm/c under the influence of the mean field. At that time,
as it was found for the calculations, most fireball-like
nucleons have left the reaction zone.

The lowest estimates of the excitation energy result from
the analyses of the earlier $^{197}$Au on Cu data at 600 MeV per nucleon 
\cite{kreutz} performed with the statistical multifragmentation models 
[5-7,10,67].
%\cite{botv1,barz1,baoanli,botv2,desesquelles1}. 
There, in order to reproduce the observed multiplicities
and charge correlations of intermediate-mass fragments, 
an ensemble of equilibrated residual nuclei was needed which is
characterized by a saturation of the excitation energy at about 8 MeV
per nucleon if the Copenhagen or Moscow models are used
\cite{botv1,barz1,botv2,desesquelles1}.
With the Berlin version of the 
statistical multifragmentation model a saturation energy as low as
6 MeV per nucleon was obtained \cite{baoanli}. The systematic study of
the complex correlation between the observables and the code input 
parameters has shown, however,
that the deduced saturation energy depends sensitively on
the measured mean fragment multiplicity \cite{desesquelles1}.
The present data, exhibiting slightly higher mean multiplicities, 
therefore lead to excitation
energies that may reach up to about 12 MeV per nucleon at small $Z_{bound}$
\cite{desesquelles2,botv4}.

A method to determine the excitation energy from the experimental data 
alone, without relying on a reaction model,
was first presented by Campi {\it et al.} \cite{campi}
and applied to the earlier $^{197}$Au + Cu data. After having determined
the yields of hydrogen isotopes by extrapolating to $Z$ = 1 from the measured 
abundances for $Z \ge$ 2,
these authors estimated the energy residing in the breakup Q value and
in the kinetic energies of the final fragments. The asymptotic value of
$E_x/A$ = 23 MeV at $Z_{bound}$ = 0 
is the sum of the binding energy of 8 MeV
and the kinetic energy of 15 MeV assigned to nucleons.
In the same type of analysis with the present data 
for $^{197}$Au + $^{197}$Au at 600 MeV per nucleon, the measured neutron 
multiplicities and their kinetic energies in the projectile frame were 
taken into account \cite{pocho1}. 
Solving the balance equations for energy, mass, and charge 
with this additional
information yields a larger number of composite 
particles, correspondingly less free nucleons, and
smaller excitation energies. A maximum of $E_x/A$ = 15 MeV for
$Z_{bound} \le$ 10 was reported
in ref. \cite{pocho1}. The data points shown in fig.~\ref{exiwoe} 
are the result of a new 
analysis with the same method from which excitation energies
up to a maximum of 17 MeV per nucleon are obtained. The 
development of this method is still in progress and further small changes 
are not excluded.

The BUU estimate seems to come closest to the two experimental
determinations of the available decay energy which fall 
in between the differing estimates with the 
intranuclear cascade and the statistical multifragmentation models.
Assuming that all analyses are realistic within their own frameworks,
the observed
ordering is quite reasonable in the sense that the formation of the
equilibrated spectator in the primary reaction stages and its evolution
towards the final breakup stage may be accompanied by the emission
of fast light particles and thus by a loss of excitation energy. 
In addition, small collective contributions due to
rotation or flow will also cause differences between the measured
energies and those derived from statistical analyses of the exit
channel configurations.
The more recent results obtained with the statistical 
multifragmentation models narrow this gap but still
leave room for a substantial preequilibrium stage or a small collective
component, even though no positive evidence for either one has emerged
from the present work.
At $Z_{bound} \approx$ 40, where the fragment multiplicity reaches its
maximum, the experimentally determined excitation energies are between
8 and 10 MeV per nucleon, close to the mean binding energy of nuclei.

\section{Conclusion}
\label{Sec_5}

The systematic set of data, measured from 400 up to 1000 MeV per nucleon
incident energy,
reveals the universal nature of multi-fragment
decays of excited spectator nuclei at relativistic energies.
It suggests that the correlation of excitation energy and 
mass of the produced spectator systems and the statistical nature
of their decay remain virtually
unchanged over the studied range of bombarding energies. In reactions with
the lighter targets, specific minimum values of the bombarding energy 
have to be exceeded in order to
reach the maximum fragment multiplicities and to achieve a complete
disassembly of the projectile spectator. The corresponding
lower limits are about 400, 800, and 1000
MeV per nucleon for collisions of gold projectiles with 
aluminum, carbon, and beryllium targets, respectively.

The results of calculations with the intranuclear cascade and statistical 
multifragmentation models permit a qualitative understanding of the observed
universality. The calculations as well as comparisons with
emulsion data for collisions at higher incident energies suggest that the 
identified production and fragmentation 
of excited spectator nuclei may persist up to very high 
energies with virtually invariant properties and slowly changing
cross sections. In the pure 
abrasion picture, this follows from the mutual connection between 
excitation energy and cross section via their geometric dependence on the 
impact parameter. The additional mechanism 
of spectator heating by scattered nucleons or produced
particles is not expected
to significantly modify this picture. 

The observed decoupling  
from the entrance-channel dynamics 
strongly suggests that the multi-fragment decay of the 
spectator occurs after it has reached statistical equilibrium.
The isotropy in the rest frame of the spectator 
and the $Z$ dependence of the intrinsic fragment momenta 
indicate that this includes the kinetic degrees of freedom.
Statistical interpretations were shown to be applicable,
and the fragmentation seems to be mainly governed by the deposited
excitation energy. The fragment
multiplicities rise with excitation energy until a maximum is reached
at about 8 to 10 MeV per nucleon, as experimentally determined for
reactions with Au projectiles.
For the more violent collisions in the regime of decreasing
fragment multiplicities, the estimates of the energy deposited in
the decaying spectator system
disagree more widely, thus leaving room for a possibly important 
role of preequilibrium emission.
This problem will need further attention if the study of equilibrium 
properties of nuclear matter is to be extended to these very high 
excitation energies.
\vspace{0.5cm}

{\it
The authors wish to thank the staff at SIS and GSI for the
excellent working conditions and J. L\"uhning and 
W. Quick for the technical support provided. Useful discussions with
A.S.~Botvina and H.C.~Britt are gratefully acknowledged.
J.P. and M.B. acknowledge the financial support 
of the Deutsche Forschungsgemeinschaft under the Contract No. Po 256/2-1 
and Be1634/1-1, respectively.
W.G.L. and L.G.S.
acknowledge the receipt of U.S. Presidential
Young Investigators Awards.
This work was supported by the European Community under
contract ERBCHGE-CT92-0003 and ERBCIPD-CT94-0091,
by the National Science
Foundation under Grants No. PHY-90-15255 and
No. PHY-92-14992, and by the U.S. Department of Energy under
Contract No. DE-FG02-87ER-40316.}

%{\bf missing:} acknowledgement of discussions:\\
%Botvina, gross, Friedman, Noerenberg


\newpage

\begin{thebibliography}{99}

\bibitem[a]{AAA}
Present Address:
Chalk River Laboratories, 
Chalk River, Ontario K0J 1J0, Canada

\bibitem[b]{BBB}
On leave from the Comision Nacional Energia Atomica, Argentina

\bibitem[c]{CCC}
Present Address:
Department of Chemistry, Indiana University, 
Bloomington, IN 47405, USA

\bibitem[d]{DDD}
Present address:
National Superconducting Cyclotron Laboratory, Michigan State University, 
East Lansing, MI 48824, USA

\bibitem[e]{JJJ}
Present address: Nuclear Science Division, Lawrence Berkeley Laboratory,
Berkeley, CA 94720, USA

\bibitem[f]{FFF}
Present address:
Department of Physics, Massachusetts Institute of Technology,
Cambridge, MA 02139, USA

\bibitem[g]{GGG}
Present address: 
Physics Department, Hope College,
Holland, MI 49223, USA

\bibitem[h]{III}
Present address: 
Max-Planck-Institut f\"ur Kernphysik,
D-69117 Heidelberg, Germany

\bibitem{ogilvie}
C.A.~Ogilvie,
J.C.~Adloff, M.~Begemann--Blaich, P.~Bouissou,
J.~Hubele, G.~Imme, I.~Iori, P.~Kreutz,
G.J.~Kunde, S.~Leray, V.~Lindenstruth, Z.~Liu,
U.~Lynen, R.J.~Meijer, U.~Milkau, W.F.J.~M\"{u}ller, C.~Ng\^{o},
J.~Pochodzalla, G.~Raciti, G.~Rudolf, H.~Sann,
A.~Sch\"{u}ttauf, W.~Seidel, L.~Stuttge, W.~Trautmann, and A.~Tucholski,
Phys.~Rev.~Lett. 67 (1991) 1214

\bibitem{hubele1}
J.~Hubele, P.~Kreutz, J.C.~Adloff, M.~Begemann--Blaich, P.~Bouissou,
G.~Imme, I.~Iori, G.J.~Kunde, S.~Leray, V.~Lindenstruth, Z.~Liu,
U.~Lynen, R.J.~Meijer, U.~Milkau, A.~Moroni, W.F.J.~M\"{u}ller, C.~Ng\^{o},
C.A.~Ogilvie, J.~Pochodzalla, G.~Raciti, G.~Rudolf, H.~Sann,
A.~Sch\"{u}ttauf, W.~Seidel, L.~Stuttge, W.~Trautmann, and A.~Tucholski,
Z.~Phys. A 340 (1991) 263

\bibitem{kreutz}
P.~Kreutz, J.C.~Adloff, M.~Begemann--Blaich, P.~Bouissou,
J.~Hubele, G.~Imme, I.~Iori, 
G.J.~Kunde, S.~Leray, V.~Lindenstruth, Z.~Liu,
U.~Lynen, R.J.~Meijer, U.~Milkau, A.~Moroni, W.F.J.~M\"{u}ller, C.~Ng\^{o},
C.A.~Ogilvie, J.~Pochodzalla, G.~Raciti, G.~Rudolf, H.~Sann,
A.~Sch\"{u}ttauf, W.~Seidel, L.~Stuttge, W.~Trautmann, and A.~Tucholski,
Nucl.~Phys. A556 (1993) 672

\bibitem{hubele2}
J.~Hubele, P.~Kreutz, V.~Lindenstruth, J.C.~Adloff, M.~Begemann--Blaich, 
P.~Bouissou, G.~Imme, I.~Iori, G.J.~Kunde, S.~Leray, Z.~Liu,
U.~Lynen, R.J.~Meijer, U.~Milkau, A.~Moroni, W.F.J.~M\"{u}ller, C.~Ng\^{o},
C.A.~Ogilvie, J.~Pochodzalla, G.~Raciti, G.~Rudolf, H.~Sann,
A.~Sch\"{u}ttauf, W.~Seidel, L.~Stuttge, W.~Trautmann, A.~Tucholski,
R.~Heck, A.R.~DeAngelis, D.H.E.~Gross, H.R.~Jaqaman, H.W.~Barz, 
H.~Schulz, W.A.~Friedman, and R.J.~Charity,
Phys.~Rev. C46 (1992) R1577

\bibitem{botv1}
A.S.~Botvina and I.N.~Mishustin, Phys.~Lett. B294 (1992) 23

\bibitem{barz1}
H.W.~Barz, W.~Bauer, J.P.~Bondorf, A.S.~Botvina, R.~Donangelo, 
H.~Schulz, and K.~Sneppen, 
Nucl.~Phys. A561 (1993) 466

\bibitem{baoanli}
Bao-An~Li, A.R.~DeAngelis, and D.H.E.~Gross, Phys.~Lett. B303 (1993) 225

\bibitem{konopka}
J.~Konopka, G.~Peilert, H.~St\"ocker, and W.~Greiner,
Prog. Part. Nucl. Phys. 30 (1993) 301

\bibitem{zheng}
Zheng Yu-Ming, Wang Fei, Sa Ben-Hao, and Zhang Xiao-Ze, 
Phys. Rev. C53 (1996) 1868

\bibitem{botv2}
A.S.~Botvina, I.N.~Mishustin, M.~Begemann--Blaich, J.~Hubele, G.~Imme, 
I.~Iori, P.~Kreutz, G.J.~Kunde, W.D.~Kunze, V.~Lindenstruth, U.~Lynen, 
A.~Moroni, W.F.J.~M\"{u}ller, C.A.~Ogilvie, J.~Pochodzalla, G.~Raciti, 
Th.~Rubehn, H.~Sann, A.~Sch\"{u}ttauf, W.~Seidel, W.~Trautmann, 
and A.~W\"orner, 
Nucl.~Phys. A584 (1995) 737

\bibitem{garcia}
J.B. Garcia and C. Cerruti,
Nucl.~Phys. A578 (1994) 597

\bibitem{botet}
R. Botet and M. Ploszajczak,
Phys. Lett. B312 (1993) 30;
Acta Physica Polonica B25 (1994) 353

\bibitem{leray}
S. Leray and S. Souza,
Proceedings of Second European Biennial Conference
on Nuclear Physics, Meg\`{e}ve 1993, edited by D. Guinet
(World Scientific, Singapore, 1995) p. 81

\bibitem{richert}
B. Elattari, J. Richert, P. Wagner, and Y.M. Zheng, 
Phys. Lett. B356 (1995) 181;
Nucl. Phys. A592 (1995) 385

\bibitem{moretto}
For a recent review see L.G.~Moretto and G.J. Wozniak,
Ann. Rev. Nucl. Part. Science 43 (1993) 379

\bibitem{gilkes}
M.L.~Gilkes, S.~Albergo, F.~Bieser, F.P.~Brady, Z.~Caccia, D.A.~Cebra,
A.D.~Chacon, J.L.~Chance, Y.~Choi, S.~Costa, J.B.~Elliott, J.A.~Hauger,
A.S.~Hirsch, E.L.~Hjort, A.~Insolia, M.~Justice, D.~Keane, J.C.~Kintner,
V.~Lindenstruth, M.A.~Lisa, U.~Lynen, H.S.~Matis, M.~McMahan, 
C.~McParland, W.F.J.~M\"{u}ller, D.L.~Olson, M.D.~Partlan, N.T.~Porile,
R.~Potenza, G.~Rai, J.~Rasmussen, H.G.~Ritter, J.~Romanski, J.L.~Romero,
G.V.~Russo, H.~Sann, R.~Scharenberg, A.~Scott, Y.~Shao, B.K.~Srivastava,
T.J.M.~Symons, M.~Tincknell, C.~Tuv\'{e}, S.~Wang, P.~Warren,
H.H.~Wieman, and K.~Wolf,
Phys.~Rev.~Lett. 73 (1994) 1590

\bibitem{pocho1}
J.~Pochodzalla, T.~M\"ohlenkamp, T.~Rubehn, A.~Sch\"{u}ttauf,
A.~W\"orner, E.~Zude, M.~Begemann--Blaich, Th.~Blaich, 
H.~Emling, A.~Ferrero, C.~Gro\ss, G.~Imme, I.~Iori, G.J.~Kunde, 
W.D.~Kunze, V.~Lindenstruth, U.~Lynen, A.~Moroni, W.F.J.~M\"{u}ller, 
B.~Ocker, G.~Raciti, H.~Sann, C.~Schwarz, W.~Seidel, V.~Serfling,
J.~Stroth, W.~Trautmann, A.~Trzcinski, A.~Tucholski, G.~Verde,
and B.~Zwieglinski,
Phys.~Rev.~Lett. 75 (1995) 1040

\bibitem{rubehn1}
Th.~Rubehn, W.F.J.~M\"{u}ller, R.~Bassini, M.~Begemann--Blaich, Th.~Blaich,
A.~Ferrero, C.~Gro\ss, G.~Imme, I.~Iori, G.J.~Kunde, W.D.~Kunze, 
V.~Lindenstruth, U.~Lynen, T.~M\"ohlenkamp, L.G.~Moretto, B.~Ocker,
J.~Pochodzalla, G.~Raciti, S.~Reito, H.~Sann, A.~Sch\"{u}ttauf,
W.~Seidel, V.~Serfling, W.~Trautmann, A.~Trzcinski, G.~Verde,
A.~W\"orner, E.~Zude, and B.~Zwieglinski,
Z. Phys. A 353 (1995) 197

\bibitem{rubehn2}
Th.~Rubehn, R.~Bassini, M.~Begemann--Blaich, Th.~Blaich,
A.~Ferrero, C.~Gro\ss, G.~Imme, I.~Iori, G.J.~Kunde, W.D.~Kunze,
V.~Lindenstruth, U.~Lynen, T.~M\"ohlenkamp, L.G.~Moretto, 
W.F.J.~M\"{u}ller, B.~Ocker,
J.~Pochodzalla, G.~Raciti, S.~Reito, H.~Sann, A.~Sch\"{u}ttauf,
W.~Seidel, V.~Serfling, W.~Trautmann, A.~Trzcinski, G.~Verde,
A.~W\"orner, E.~Zude, and B.~Zwieglinski,
Phys. Rev. C53 (1996) 993

\bibitem{rubehn3}
Th.~Rubehn, R.~Bassini, M.~Begemann--Blaich, Th.~Blaich,
A.~Ferrero, C.~Gro\ss, G.~Imme, I.~Iori, G.J.~Kunde, W.D.~Kunze,
V.~Lindenstruth, U.~Lynen, T.~M\"ohlenkamp, L.G.~Moretto,
W.F.J.~M\"{u}ller, B.~Ocker,
J.~Pochodzalla, G.~Raciti, S.~Reito, H.~Sann, A.~Sch\"{u}ttauf,
W.~Seidel, V.~Serfling, W.~Trautmann, A.~Trzcinski, G.~Verde,
A.~W\"orner, E.~Zude, and B.~Zwieglinski,
Phys. Rev. C53 (1996) 3143

\bibitem{bormio_95}
W.~Trautmann {\it et al.},
Proceedings of the XXXIII International Winter Meeting on
Nuclear Physics, Bormio, 1995, 
edited by I. Iori (Ricerca Scientifica ed Educazione Permanente,
Milano, 1995) p. 372

\bibitem{desouza}
R.T.~De Souza, N.~Carlin, Y.D.~Kim, J.~Ottarson, L.~Phair, D.R.~Bowman,
C.K.~Gelbke, W.G.~Gong, W.G.~Lynch, R.A.~Pelak, T.~Peterson, G.~Poggi,
M.B.~Tsang, and H.M.~Xu,
Nucl. Instrum. Methods Phys. Res. A295 (1990) 109

\bibitem{tsang}
M.B.~Tsang, W.C.~Hsi, W.G.~Lynch, D.R.~Bowman, C.K.~Gelbke, M.A.~Lisa,
G.F.~Peaslee, G.J.~Kunde, M.L.~Begemann--Blaich, T. Hofmann, J.~Hubele, 
J.~Kempter, P.~Kreutz, W.D.~Kunze, V.~Lindenstruth, U.~Lynen, M.~Mang,
W.F.J.~M\"{u}ller, M.~Neumann, B.~Ocker, C.A.~Ogilvie, J.~Pochodzalla, 
F.~Rosenberger, H.~Sann, A.~Sch\"{u}ttauf, V.~Serfling, J.~Stroth, 
W.~Trautmann, A.~Tucholski, A.~W\"orner, E.~Zude, B.~Zwieglinski,
S.~Aiello, G.~Imme, V.~Pappalardo, G.~Raciti, R.J.~Charity,
L.G.~Sobotka, I.~Iori, A.~Moroni, R.~Scardaoni, A.~Ferrero, W.~Seidel, 
Th.~Blaich, L.~Stuttge, A.~Cosmo, W.A.~Friedman, and G.~Peilert,
Phys. Rev. Lett. 71 (1993) 1502
                                                    
\bibitem{reisdorf}
W. Reisdorf {\it et al.},
Proceedings of the International Workshop XXII, Hirschegg, 1994, edited by
H. Feldmeier and W. N\"orenberg (GSI, Darmstadt, 1994) p. 93

\bibitem{suemmerer}
K.~S\"ummerer, W.~Br\"uchle, D.J.~Morrissey, M.~Sch\"adel, B.~Szweryn,
and Yang~Weifan, 
Phys. Rev. C42 (1990) 2546

\bibitem{voli}
V.~Lindenstruth, PhD thesis, Universit\"at Frankfurt, 1993, 
report GSI-93-18

\bibitem{kunde1}
G.J.~Kunde, J.~Pochodzalla, J.~Aichelin, E.~Berdermann, B.~Berthier,
C.~Cerruti, C.K.~Gelbke, J.~Hubele, P.~Kreutz, S.~Leray, R.~Lucas,
U.~Lynen, U.~Milkau, W.F.J.~M\"{u}ller, C.~Ng\^o, C.H.~Pinkenburg,
G.~Raciti, H.~Sann, and W.~Trautmann,
Phys. Lett. B272 (1991) 202

\bibitem{barz2}
H.W. Barz, H.~Schulz, and G.F. Bertsch,
Phys. Lett. B217 (1989) 397

\bibitem{boal}
D.H. Boal, J.N. Glosli, and C. Wicentowich,
Phys. Rev. Lett. 62 (1989) 737

\bibitem{barz3}
H.W. Barz, J.P. Bondorf, R.~Donangelo, H.~Schulz, and K.~Sneppen,
Phys. Lett. B228 (1989) 453

\bibitem{bauer1}
W. Bauer,
Phys. Rev. C51 (1995) 803

\bibitem{goldhaber}
A.S. Goldhaber, 
Phys. Lett. 53B (1974) 306

\bibitem{milkau}
U.~Milkau, M.L.~Begemann--Blaich, E.-M.~Eckert, G.~Imme, P.~Kreutz,
A.~K\"uhmichel, M.~Lattuada, U.~Lynen, C.~Mazur, W.F.J.~M\"{u}ller,
J.B.~Natowitz, C.~Ng\^o, J.~Pochodzalla, G.~Raciti, M.~Ribrag,
H.~Sann, W.~Trautmann, and R.~Trockel,
Z. Phys. A346 (1993) 227

\bibitem{botv3}
A.S. Botvina and D.H.E. Gross,
Phys. Lett. B344 (1995) 6; Nucl.~Phys. A592 (1995) 257
                               
\bibitem{kunde2}
G.J.~Kunde, W.C.~Hsi, W.D.~Kunze, A.~Sch\"{u}ttauf, A.~W\"orner, 
M.~Begemann--Blaich, Th.~Blaich, D.R.~Bowman, R.J.~Charity,
A.~Cosmo, A.~Ferrero, C.K.~Gelbke, J.~Hubele, G.~Imme, I.~Iori, 
P.~Kreutz, V.~Lindenstruth, M.A.~Lisa, W.G.~Lynch, U.~Lynen,
M.~Mang, T.~M\"ohlenkamp, A.~Moroni, W.F.J.~M\"{u}ller, M.~Neumann,
B.~Ocker, C.A.~Ogilvie, G.F.~Peaslee, J.~Pochodzalla, G.~Raciti, 
T.~Rubehn, H.~Sann, W.~Seidel, V.~Serfling, L.G.~Sobotka, J.~Stroth,
L.~Stuttge, S.~Tomasevic, W.~Trautmann, M.B.~Tsang, A.~Tucholski,
G.~Verde, C.W.~Williams, E.~Zude, and B.~Zwieglinski,
Phys. Rev. Lett. 74 (1995) 38

\bibitem{wadding}
C.J. Waddington and P.S. Freier,
Phys. Rev. C31 (1985) 888

\bibitem{lips}
V.~Lips, R.~Barth, H.~Oeschler, S.P.~Avdeyev, V.A.~Karnaukhov, 
W.D.~Kuznetsov, L.A.~Petrov, O.V.~Bochkarev, L.V.~Chulkov, E.A.~Kuzmin,
W.~Karcz, W.~Neubert, and E.~Norbeck,
Phys. Rev. Lett. 72 (1994) 1604

\bibitem{kwiat2}
K.~Kwiatkowski, K.B.~Morley, E.~Renshaw Foxford, D.S.~Bracken, V.E.~Viola,
N.R.~Yoder, R.~Legrain, E.C.~Pollacco, C.~Volant, W.A.~Friedman, 
R.G.~Korteling, J.~Brzychczyk, and H.~Breuer,
Phys. Rev. Lett. 74 (1995) 3756

\bibitem{rusch}
G.~Rusch, W.~Heinrich, B.~Wiegel, E.~Winkel, and J.~Dreute,
Phys. Rev. C49 (1994) 901

\bibitem{jain}
P.L. Jain and G. Singh,
Phys. Rev. C46 (1992) R10

\bibitem{jain2}
P.L. Jain, G. Singh, and A. Mukhopadhyay,
Phys. Rev. C50 (1994) 1085

\bibitem{klmm1}
M.L.~Cherry, A.~Dabrowska, P.~Deines-Jones, A.J.~Dubinina, R.~Holynski,
W.V.~Jones, E.D.~Kolganova, A.~Olszewski, E.A.~Pozharova, K.~Sengupta,
T.Yu.~Skorodko, V.A.~Smirnitski, M.~Szarska, C.J.~Waddington, J.P.~Wefel,
B.~Wilczynska, and W.~Wolter,
Z.~Phys. C 63 (1994) 549

\bibitem{klmm2}
M.L.~Cherry, A.~Dabrowska, P.~Deines-Jones, R.~Holynski,
W.V.~Jones, E.D.~Kolganova, A.~Olszewski, K.~Sengupta,
T.Yu.~Skorodko, M.~Szarska, C.J.~Waddington, J.P.~Wefel,
B.~Wilczynska, B.~Wosiek, and W.~Wolter,
Phys. Rev. C 52 (1995) 2652

\bibitem{peaslee}
G.F.~Peaslee, M.B.~Tsang, C.~Schwarz, M.J.~Huang, W.S.~Huang,
W.C.~Hsi, C.~Williams, W.~Bauer, D.R.~Bowman, M.~Chartier,
J.~Dinius, C.K.~Gelbke, T.~Glasmacher, D.O.~Handzy, M.A.~Lisa,
W.G.~Lynch, C.M.~Mader, L.~Phair, M-C.~Lemaire, S.R.~Souza,
G.~Van Buren, R.J.~Charity, L.G.~Sobotka, G.J.~Kunde, 
U.~Lynen, J.~Pochodzalla, H.~Sann, W.~Trautmann, D.~Fox, R.T.~de Souza,
G.~Peilert, W.A.~Friedman, and N.~Carlin,
Phys. Rev. C49 (1994) R2271

\bibitem{pienko}
L.~Pienkowski, H.G.~Bohlen, J.~Cugnon, H.~Fuchs, J.~Galin, B.~Gatty,
B.~Gebauer, D.~Guerreau, D.~Hilscher, D.~Jacquet, U.~Jahnke, M.~Josset,
X.~Ledoux, S.~Leray, B.~Lott, M.~Morjean, A.~P\'{e}ghaire, G.~R\"oschert,
H.~Rossner, R.H.~Siemssen, and C.~St\'{e}phan,
Phys. Lett. B336 (1994) 147

\bibitem{dreute}
J.~Dreute, W.~Heinrich, G.~Rusch, and B.~Wiegel,
Phys. Rev. C44 (1991) 1057

\bibitem{limit1}
S.B. Kaufman and E.P. Steinberg,
Phys. Rev. C22 (1980) 167

\bibitem{limit2}
R.E.L. Green and R.G. Korteling,
Phys. Rev. C22 (1980) 1594

\bibitem{porile}
N.T.~Porile, A.J.~Bujak, D.D.~Carmony, Y.H.~Chung, L.J.~Gutay, A.S.~Hirsch,
M.~Mahi, G.L.~Paderewski, T.C.~Sangster, R.P.~Scharenberg, 
and B.C.~Stringfellow,
Phys. Rev. C39 (1989) 1914

\bibitem{berthier}
B.~Berthier, R.~Boisgard, J.~Julien, J.M.~Hisleur, R.~Lucas, C.~Mazur,
C.~Ng\^o, M.~Ribrag, and C.~Cerruti,
Phys. Lett. B193 (1987) 417

\bibitem{traut}
W.~Trautmann, U.~Milkau, U.~Lynen, and J.~Pochodzalla,
Z.~Phys. A 344 (1993) 447

\bibitem{gaimard}
J.-J. Gaimard and K.-H.~Schmidt,
Nucl. Phys. A531 (1991) 709

\bibitem{brohm}
K.-H.~Schmidt, T.~Brohm, H.-G.~Clerc, M.~Dornik, M.~Fauerbach, H.~Geissel,
A.~Grewe, E.~Hanelt, A.~Junghans, A.~Magel, W.~Morawek, G.~M\"unzenberg,
F.~Nickel, M.~Pf\"utzner, C.~Scheidenberger, K.~S\"ummerer, D.~Vieira,
B.~Voss, and C.~Ziegler,
Phys. Lett. B300 (1993) 313;\\
T.~Brohm and K.-H.~Schmidt,
Nucl. Phys. A569 (1994) 821

\bibitem{kno}
Z. Koba, H.B. Nielsen, and P.~Olesen,
Nucl. Phys. B40 (1972) 317

\bibitem{ziping}
For a recent review see Ziping Chen,
Int. J. Mod. Phys. E2 (1993) 285

\bibitem{yariv}
Y.~Yariv and Z.~Fraenkel,
Phys. Rev. C20 (1979) 2227;
Phys. Rev. C24 (1981) 488

\bibitem{cugnon}
J.~Cugnon, 
Nucl. Phys. A462 (1987) 751

\bibitem{brown}
G.E.~Brown, 
Nucl. Phys. A488 (1988) 689c
          
\bibitem{kwiat1}
K.~Kwiatkowski, W.A.~Friedman, L.W.~Woo, V.E.~Viola, E.C.~Pollacco, 
C.~Volant, and S.J. Yennello,
Phys. Rev. C49 (1994) 1516

\bibitem{harris}
J.W.~Harris, G.~Odyniec, H.G.~Pugh, L.S.~Schroeder, M.L.~Tincknell,
W.~Rauch, R.~Stock, R.~Bock, R.~Brockmann, A.~Sandoval, H.~Str\"obele,
R.E.~Renfordt, D.~Schall, D.~Bangert, J.P.~Sullivan, K.L.~Wolf, 
A.~Dacal, C.~Guerra, and M.E.~Ortiz,
Phys. Rev. Lett. 58 (1987) 463

\bibitem{muentz}
C.~M\"untz, P.~Baltes, H.~Oeschler, A.~Sartorius, A.~Wagner, 
W.~Ahner, R.~Barth, M.~Cie\'slak, M.~Debowski, E.~Grosse, W.~Henning, 
P.~Koczo\'n, D.~Mi\'skowiec, R.~Schicker, P.~Senger, C.~Bormann,
D.~Brill, Y.~Shin, J.~Stein, R.~Stock, H.~Str\"obele, B.~Kohlmeyer, 
H.~P\"oppl, F.~P\"uhlhofer, J.~Speer, K.~V\"olkel, and W.~Walus,
Z. Phys. A352 (1995) 175, and references therein

\bibitem{toneev}
V.S.~Barashenkov and V.D.~Toneev,
Interactions of high-energy particles and atomic nuclei with nuclei,
Moscow, Atomizdat, 1972 (in Russian);\\
V.D.~Toneev, private communication (1994)

\bibitem{deangelis}
D.H.E.~Gross, Rep.~Prog.~Phys.~ 53 (1990) 605;\\
A.R.~DeAngelis and D.H.E.~Gross,
Comput. Phys. Commun. 76 (1993) 113

\bibitem{friedman}
W.A. Friedman,
Phys. Rev. C42 (1990) 667

\bibitem{papp}
G.~Papp and W.~N\"orenberg, 
preprint GSI-95-30 (1995)

\bibitem{bauer2}
W. Bauer,
private communication (1991)

\bibitem{desesquelles1}
P. D\'esesquelles, J.P.~Bondorf, I.N.~Mishustin, and A.S.~Botvina,
Nucl. Phys. A, in print

\bibitem{desesquelles2}
P. D\'esesquelles,
private communication (1996)

\bibitem{botv4}
A.S. Botvina and Hongfei Xi,
private communication (1996)

\bibitem{campi}
X.~Campi, H.~Krivine, and E.~Plagnol,
Phys. Rev. C50 (1994) R2680

\end{thebibliography}
\end{document}